\documentclass[prd, aps, nofootinbib, preprintnumbers, showpacs, superscriptaddress, twocolumn]{revtex4}

\usepackage{amssymb}
\usepackage[T1]{fontenc}
\usepackage[english]{babel}
\usepackage{graphicx}
\usepackage[latin9]{inputenc}
\usepackage{mathpazo}

\newcommand{\be}{\begin{equation}}
\newcommand{\ee}{\end{equation}}

\begin{document}

\title{Cauchy--perturbative matching revisited: tests in spherical
symmetry}

\author{Burkhard        Zink}        \email{bzink@mpa-garching.mpg.de}
\affiliation{Max-Planck-Institut           f\"ur          Astrophysik,
Karl-Schwarzschild-Str.\ 1, 85741 Garching bei M\"unchen, Germany}

\author{Enrique           Pazos}           \email{enrique@cct.lsu.edu}
\affiliation{Department of Physics  and Astronomy, 202 Nicholson Hall,
Louisana State University, Baton Rouge, LA 70803, USA}

\affiliation{Center for Computation and Technology, 302 Johnston Hall,
  Louisana State University, Baton Rouge, LA 70803, USA}

\affiliation{Departamento de  Matem\'atica, Universidad de  San Carlos
  de Guatemala}

\author{Peter Diener} \email{diener@cct.lsu.edu}
\homepage{http://www.cct.lsu.edu/}

\affiliation{Department of Physics  and Astronomy, 202 Nicholson Hall,
  Louisana State University, Baton Rouge, LA 70803, USA}

\affiliation{Center for Computation and Technology, 302 Johnston Hall,
  Louisana State University, Baton Rouge, LA 70803, USA}

\author{Manuel Tiglio} \email{tiglio@cct.lsu.edu}
\homepage{http://www.cct.lsu.edu/}

\affiliation{Department of Physics  and Astronomy, 202 Nicholson Hall,
  Louisana State University, Baton Rouge, LA 70803, USA}

\affiliation{Center for Computation and Technology, 302 Johnston Hall,
  Louisana State University, Baton Rouge, LA 70803, USA}

\begin{abstract}
During the  last few  years progress has  been made on  several fronts
making it  possible to revisit Cauchy--perturbative  matching (CPM) in
numerical relativity in a more robust and accurate way.  This paper is
the first  in a series where  we plan to  analyze CPM in the  light of
these new results.

One  of the  new developments  is an  understanding of  how  to impose
constraint-preserving boundary  conditions (CPBC); though  most of the
related research has been driven by outer boundaries, one can use them
for matching interface boundaries as well. Another front is related to
numerically  stable evolutions  using multiple  patches, which  in the
context  of CPM allows  the matching  to be  performed on  a spherical
surface, thus avoiding  interpolations between Cartesian and spherical
grids. One  way of achieving  stability for such schemes  of arbitrary
high  order is  through the  use  of penalty  techniques and  discrete
derivatives satisfying summation by  parts (SBP).  Recently, new, very
efficient  and  high order  accurate  derivatives  satisfying SBP  and
associated dissipation operators have been constructed.

Here we  start by  testing all  these techniques applied  to CPM  in a
setting that  is simple enough to  study all the  ingredients in great
detail: Einstein's equations in spherical symmetry, describing a black
hole  coupled to  a  massless scalar  field.   We show  that with  the
techniques    described    above,    the    errors    introduced    by
Cauchy--perturbative matching are very  small, and that very long term
and accurate  CPM evolutions can  be achieved.  Our tests  include the
accretion and ring-down phase of  a Schwarzschild black hole with CPM,
where  we find  that the  discrete  evolution introduces,  with a  low
spatial resolution of $\Delta r =  M/10$, an error of $0.3\%$ after an
evolution time of  $1,000,000 \, M$.  For a black  hole of solar mass,
this corresponds  to approximately $5 \,  s$, and is  therefore at the
lower  end of  timescales discussed  e.g.  in  the collapsar  model of
gamma-ray burst engines.

\end{abstract}

\pacs{04.25.Dm, 04.25.Nx, 04.70.Bw}

\maketitle

\section{Introduction}

It is generally  expected that the geometry of  compact sources should
resemble flat spacetime  at large enough distances.  This  is true not
only qualitatively,  but through very precise  falloff conditions that
are built  into the formal definition of  asymptotic flatness.  Within
this definition, the deviations from flat spacetime are well described
(in the sense of the leading order behaviour of an expansion in powers
of  ``$1/r$'')   by  perturbations  of   the  Schwarzschild  spacetime
\cite{Misner73}.

Such perturbations can in turn  be studied through the gauge invariant
Regge-Wheeler and Zerilli  (RWZ) formalisms \cite{Regge57, Zerilli70}.
These allow  one to derive,  after a spherical  harmonic decomposition
(that is, for each ``$(\ell,m)$''), two master evolution equations for
the  truly gauge  invariant, linearized  physical degrees  of freedom.
Due to  the multipole decomposition, these equations  involve only one
spatial  coordinate  (the  radial   one).   The  fact  that  they  are
one-dimensional implies that these  master equations can be solved for
very  large  computational  domains  with  very  modest  computational
resources.  On the other hand, three-dimensional Cauchy codes are very
demanding on their resource  requirements. Even though mesh refinement
can help in this respect, there is a limit to how much one can coarsen
the grid in the asymptotic region; this limit is set by the resolution
required  to reasonably  represent wave  propagation in  the radiative
zone. The  use of  a grid structure  adapted to the  physical geometry
(possibly through  multiple patches) can  also help \cite{Lehner2005a,
Thornburg2004:multipatch-BH-excision,  Kidder01a}, but one  still ends
up  imposing  artificial   (even  if  constraint-preserving)  boundary
conditions at the outer boundary.   For example, one in general misses
information about the geometry outside the domain \cite{Lau:2005}.

Two  approaches  that  at  the  same  time  provide  wave  extraction,
physically motivated boundary conditions, and extend the computational
domain   to   the    radiative   regime   are   Cauchy--characteristic
\cite{Winicour05,  Bishop96} and  Cauchy--perturbative  matching (CPM)
\cite{Rezzolla99a, Rupright98, Gomez98a}; this paper is concerned with
the latter. The  idea is to match at each  timestep a fully non-linear
Cauchy   code    to   an   outer    one   solving,   say,    the   RWZ
equations\footnote{Even though  including the angular  momentum of the
background is a high order correction in terms of powers of $1/r$, one
might, in principle, try to  solve for perturbations of Kerr spacetime
(as opposed to Schwarzschild).}.

This paper is the  first one in a series where we  plan to revisit CPM
in  the  light  of  some  recent technical  developments  ---which  we
describe  below--- that  should  help in  its implementation.   Before
discussing  these  developments,  we  point  out  and  summarize  some
features present in  the original implementation of CPM  which we hope
to improve on:
\begin{enumerate}
\item  The non-linear  Cauchy equations  were solved  on  a Cartesian,
cubic  grid.   On the  other  hand, the  RWZ  equations  use a  radial
coordinate  for the spatial  dimension.  Mixing  Cartesian coordinates
with spherical ones leads to the need for interpolation back and forth
between both  grids.  Especially when  using high order  methods, this
type of interpolation  might not only be complicated  but also subtle:
depending on how it is done  it might introduce noise and sometimes it
might even be a source of numerical instabilities.

\item When injecting  data from the perturbative module  to the Cauchy
code  and vice  versa boundary  conditions  were given  to all  modes,
irrespectively  of their  propagation  speed and  without taking  into
account  the existence  of constraint  violating boundary  modes.  One
would intuitively expect a cleaner matching if boundary conditions are
given  according to  the  characteristic (propagation)  speeds of  the
different  modes,  and  even  cleaner  if  constraint-preservation  is
automatically built in during the matching.

\item Low  order numerical schemes, which result  in slow convergence,
were used.
\end{enumerate}

In recent years there has been progress on several related fronts that
should in  principle help in  the implementation of CPM.   We describe
these new results  next\footnote{There is actually another ingredient:
the use  of a generalized  perturbative formalism that allows  for any
(spherically symmetric) slicing of the background Schwarzschild metric
\cite{sarbach:2001qq}. However, since  such ingredient will not appear
in the  simplified model that  we look at  in this paper, we  skip its
discussion here.}:

\begin{enumerate}
\item The  first improvement  is the ability  to implement  smooth (in
particular,   spherical)   boundaries    in   3D   Cauchy   evolutions
\cite{Lehner2005a,  Thornburg2004:multipatch-BH-excision,  Kidder01a}.
One important advantage  of this is the fact that  the matching can be
performed  ---to  either  a  perturbative or  a  characteristic  outer
module---  without the  need for  interpolation between  spherical and
Cartesian  grids.  In  that  way a  possible  source of  noise can  be
eliminated.  It is now understood how to match different domains using
schemes  of arbitrary  high  order  while at  the  same time  ensuring
numerical  stability.  One  way  of doing  so  is through  the use  of
multiple patches  (much in  the same way  multiple charts are  used in
differential  geometry),   penalty  terms  and   difference  operators
satisfying  summation  by  parts~\cite{Lehner2005a} (more  about  this
below). This is  the approach we shall explore here  in the context of
CPM\footnote{Regardless of whether matching is present or not, the use
of  multiple coordinate  patches has  advantages when  modelling black
holes  through  excision of  the  singularity  from the  computational
domain.}.

\item    The    second   improvement    is    the   construction    of
constraint-preserving boundary conditions (CPBC). Several efforts have
by  now  reported numerically  stable  (in  the  sense of  convergent)
implementations   of   such   boundary   conditions  for   the   fully
three-dimensional  non-linear  Einstein's  equations  \cite{Sarbach04,
Kidder2005:boundary-conditions, Szilagyi02a}.  Furthermore, there have
been reports  in the  context of Cauchy--characteristic  matching that
significant  improvements  are obtained  when  this  type of  boundary
conditions are used in  the matching~\cite{Szilagyi02a}.  With this in
mind, we will test their use in CPM.

\item  Lastly,  new,  accurate  and efficient  high  order  difference
operators  satisfying SBP  and associated  dissipative  operators have
been     constructed    recently     \cite{Strand1994a,    Strand1996,
Mattsson2004a,  Diener05b}.  As mentioned  above, in  conjunction with
certain penalty interface treatment such operators guarantee numerical
stability when ``glueing'' together different computational grids.  We
will test these operators in the context of CPM.
\end{enumerate}

We    have   incorporated    these   techniques,    i.e.,   high-order
summation-by-parts  finite  differencing  and  dissipation  operators,
multiple     coordinate    patches     with     penalty    inter-patch
constraint-preserving  boundary  conditions  and  Cauchy--perturbative
matching,  into a  spherically symmetric  numerical code  evolving the
Einstein--Christoffel form  of the field  equations \cite{Anderson99},
minimally coupled  to a  Klein-Gordon field. Using  this tool,  we can
test the  performance of the  numerical methods in a  non-trivial, but
easily reproducible and  computationally inexpensive setting, and gain
experience   for  three-dimensional   applications.    The  evolutions
presented  here model black  holes with  excision in  isolation, under
dynamical  slicings, and  black holes  accreting scalar  field pulses,
which are used as a scalar analogue of gravitational radiation.

The plan of this paper  is as follows.  In section \ref{sec:equations}
we introduce the continuum system and the numerical techniques we have
used.   Results are  presented in  section \ref{sec:results},  where a
black  hole  is  evolved  successively  from  simple  settings,  i.e.,
single-patch, isolated, Killing-field adapted gauges, to more involved
ones   including  Cauchy--perturbative   matching  and   scalar  pulse
accretion.   Finally,   in  section  \ref{sec:conclusions},   we  draw
conclusions and give an outlook to future work.

\section{Equations and Methods}
\label{sec:equations}

\subsection{Evolution equations and constraint-preserving boundary conditions}

In   this  paper   we  use   the  Einstein--Christoffel   (EC)  system
\cite{Anderson99} in  spherical symmetry.   We follow the  notation of
Ref.\ \cite{Kidder00a}; in particular, the densitized lapse is denoted
by  $\alpha   =  N  g^{-1/2}$,   and  $\tilde{\alpha}  =   \alpha  r^2
\sin(\theta)$  is  introduced  for  convenience.   Here,  $g$  is  the
determinant  of the  3-metric and  $N$ the  lapse function,  while the
4-metric is written as
\begin{eqnarray}
ds^2 =  -N^2 dt^2 + g_{rr}  (dr + \beta  dt)^2 + r^2 g_T  (d\theta^2 +
\sin^2 \theta d\phi^2) \nonumber
\end{eqnarray}

The vacuum part  of the evolution equations in  spherical symmetry for
this formulation constitute a symmetric hyperbolic system of six first
order differential equations. The  vacuum variables are the two metric
and extrinsic curvature components
$$
g_{rr},\ g_{T},\ K_{rr},\ K_T,
$$
where the extrinsic curvature is written as
$$
K_{ij} = K_{rr}dr^2 + r^2 K_T (d\theta^2 + \sin^2 \theta d\phi^2),
$$
plus  two auxiliary variables  needed to  make Einstein's  equations a
first order system. These extra variables are defined as
\begin{eqnarray}
f_{rrr} &=& \frac{g'_{rr}}{2} + 4 \frac{g_{rr} f_{rT}}{g_T}, \nonumber
\\ f_{rT} &=& \frac{g'_T}{2} + \frac{g_T}{r}. \nonumber
\end{eqnarray} 

In addition, a massless Klein-Gordon field is minimally coupled to the
geometry \cite{Kidder01a, Calabrese02d}. The scalar field equation
\begin{eqnarray}
g^{ab} \nabla_a\nabla_b \Psi = 0 \nonumber
\end{eqnarray}
is  converted  into  a  first  order system  by  introduction  of  the
variables
\begin{eqnarray}
\Pi &=&  \frac{1}{N}(\beta \Psi' - \dot{\Psi}), \nonumber  \\ \Phi &=&
\Psi'. \nonumber
\end{eqnarray}
Throughout  this  paper  the   'prime'  and  'dot'  represent  partial
derivatives with respect to $r$ and $t$, respectively.

Constraint preserving boundary conditions are imposed by analyzing the
characteristic modes of the  main and constraint evolution systems, as
discussed  in \cite{Calabrese02d}.  These  modes and  their associated
characteristic  speeds are  summarized  in Table~\ref{ta:modes}.   For
illustration purposes,  we also show  the direction of  propagation of
each  mode  in the  Schwarzschild  spacetime in  Painlev\'e-Gullstrand
coordinates \cite{Painleve1921, Gullstrand:1922, Martel:2000rn}.

\begin{table}
\centering
\begin{tabular}{l|c|c|c}
\hline \hline Mode  & Speed & r<2M  & r>2M \\ \hline $u_1  = g_{rr}$ &
$\beta$ & left & left \\ $u_2 = g_T$ & $\beta$ & left & left \\ $u_3 =
K_{rr} - f_{rrr} \, g_{rr}^{-1/2}$ & $\beta + \tilde{\alpha} \, g_T$ &
left  & left  \\ $u_4  = K_T  - f_{rT}  \, g_{rr}^{-1/2}$  &  $\beta +
\tilde{\alpha}  g_T$ &  left &  left  \\ $u_5  = K_{rr}  + f_{rrr}  \,
g_{rr}^{-1/2}$ & $\beta - \tilde{\alpha} g_T$ & left & right \\ $u_6 =
K_T + f_{rT} \, g_{rr}^{-1/2}$ & $\beta - \tilde{\alpha} g_T$ & left &
right \\ $u_7 = \Pi + \Phi \, g_{rr}^{-1/2}$ & $\beta - \tilde{\alpha}
g_T$ & left & right \\ $u_8  = \Pi - \Phi \, g_{rr}^{-1/2}$ & $\beta +
\tilde{\alpha} g_T$ & left & left \\ \hline \hline
\end{tabular}
\caption{Characteristic  modes   for  Einstein-Christoffel  system  in
spherical  symmetry,   and  their  direction  of   propagation  for  a
Schwarzschild  spacetime  in  Painlev\'e-Gullstrand  coordinates  with
respect to  the vector field  $\partial_r$.  In this gauge,  all modes
are outflow at the inner boundary,  if it is located at $r< 2M$, while
boundary conditions  have to  be applied to  the incoming  modes $u_1,
u_2, u_3, u_4$ and $u_8$ at the outer boundary, assuming is is located
at $r >2 M$.}
\label{ta:modes}
\end{table}
From  Table~\ref{ta:modes}  we   notice  that  for  the  Schwarzschild
spacetime there are four  ingoing and two outgoing gravitational modes
at the outer boundary, and therefore expect the same count to hold for
perturbations  thereof.  Boundary  conditions for  the  incoming modes
$u_1, u_2$ and  $u_4$ are fixed by the CPBC  procedure. Thus, the only
free  incoming modes  are $u_3$,  which  represents a  gauge mode  and
$u_8$, which  represents a  physical one (see  \cite{Calabrese02d} for
more details). Boundary conditions do  not need to be specified at the
inner boundary if it is  located inside the event horizon, because all
modes are outflow then.

\subsection{Cauchy--perturbative matching}

Since there is no radiative degree of freedom in spherically symmetric
spacetimes,  we  use  the  massless  Klein-Gordon field  as  a  scalar
analogue   of  gravitational   waves.    To  emulate   the  setup   of
three-dimensional   Cauchy--perturbative   matching   as  closely   as
possible,  the  scalar  wave  is  evolved  on  a  fixed  Schwarzschild
background in a ``perturbative'' patch defined for $r\ge {r_m}$, while
the  fully   non-linear  Einstein's  equations  are   evolved  in  the
``Cauchy'' patch,  defined for $r\in[r_e,r_m]$, where  $r_e$ and $r_m$
denotes the excision radius and the matching radius, respectively.

The  fact  that  we are  using  CPBC  allows  us  to perform  a  clean
matching. From the analytical point  of view our matching works in the
following way: As mentioned above,  after the CPBC procedure, only two
free characteristic modes are entering the Cauchy computational domain
(at  $r=r_m$), denoted by  $u_3$ and  $u_8$. Since  in a  very precise
sense $u_3$ is  a gauge mode, we are free  to give boundary conditions
to  it  in  a  very  simple  way:  we  just  set  it  to  its  initial
value. Regarding $u_8$,  we use the "perturbative'' value  of the same
quantity  coming  from the  perturtative  domain  as counterpart,  and
communicate these two  modes (how this is done  at the numerical level
is explained below).  Similarly, there is only one characteristic mode
entering the  perturbative domain, which is the  linearized version of
$u_7$. We therefore communicate  the non-linear and linear versions of
that mode as well.

\subsection{Discrete techniques}
\label{sec:sbp}

Given a well-posed initial-boundary value problem for Einstein's field
equations, we construct a stable and accurate discrete system by using
operators satisfying the SBP  property.  In short, a finite difference
operator,  $D$,  satisfies  SBP  on  a  computational  domain  $[a,b]$
discretized using grid points $i=1,\ldots,n$ and a grid spacing $h$ if
\begin{equation}
\langle u,Dv \rangle + \langle v,Du \rangle = \left(uv\right)|_a^b
\end{equation}
holds for all grid functions  $u,v$. Here the scalar product, $\Sigma$
is defined in terms of its coefficients $\sigma_{ij}$ by
\begin{equation}
\langle u,v\rangle = h\sum_{i,j=1}^n u_i v_j \sigma_{ij}. \label{prod}
\end{equation}

In this paper  we use the new, efficient, and  accurate high order SBP
difference operators and  associated dissipation operators constructed
in Ref.\ \cite{Diener05b}.  Thus, as mentioned, this paper also serves
as an extra test of those new operators.

SBP operators are standard centered finite difference operators in the
interior of the  domain, but the stencils are  modified to yield lower
order operators in  a region close to the  boundaries (at the boundary
itself the stencil  is completely one sided). There  are several types
of SBP operators depending on the properties of the norm. The simplest
are the diagonal  norm operators. They have the  advantage that SBP is
guaranteed to  hold in  several dimensions by  simply applying  the 1D
operator  along each  direction and  that numerical  stability  can be
guaranteed by discrete energy estimates  in a wide range of cases. The
main disadvantage  is that the order  of the operator at  and close to
the  boundary is  only  half the  interior  order. We  denote the  SBP
operators by  the interior  and boundary order  and consider  here the
diagonal operators $D_{2-1}$,  $D_{4-2}$, $D_{6-3}$ and $D_{8-4}$. The
second type is  the restricted full norm operators,  where the norm is
diagonal  at  the  boundary  but  has  a  non-diagonal  block  in  the
interior. The  advantage of these operators  is that the  order at and
close to  the boundary is only  one order lower than  in the interior,
while the disadvantage is that schemes based on these operators may be
unstable without the use of dissipation. The restricted full operators
we use here are $D_{4-3}$ and $D_{6-5}$.

If  the  computational  domain   is  split  into  several  sub-domains
(``patches''), the discrete representation requires a stable technique
to communicate the solution at  inter-patch boundaries. We make use of
a                            penalty                            method
\cite{Lehner-Reula-Tiglio-2004:multipatch-scalar-field-Kerr-background},
which adds  a damping  term to  the right hand  side of  the evolution
equation  at  the  boundary  point  in  a  way  which  retains  linear
stability.   The  method has  a  free  parameter,  called $\delta$  in
Ref. \cite{Lehner-Reula-Tiglio-2004:multipatch-scalar-field-Kerr-background},
which determines how much the difference between characteristic fields
on  either side of  the inter-patch  boundary is  penalized. Different
values of $\delta$ result in different amount of energy dissipation at
the inter-patch  boundary and  can in principle  be chosen so  that no
energy is dissipated (this is marginally stable). Usually the value of
$\delta$ is chosen  such that some dissipation of  energy occurs. With
constant values  of $\delta$ the amount of  dissipation decreases with
resolution.

\subsection{Numerical code}

        For the  purposes of this paper, a  one-dimensional code which
supports constraint-preserving boundaries,  multiple grid patches, and
the  use   of  the  aforementioned  high  order   SBP  derivative  and
dissipation operators  has been developed.   In addition, the  code is
able  to  reproduce  the   (single  grid  and  without  CPM  matching)
second-order methods of  Ref.\ \cite{Calabrese02d} for comparison.  We
use the methods  of lines, and the time integration  is performed by a
4th order Runge-Kutta method.  The  grid patches that we consider here
are  not intersecting,  but touching.   This implies,  that  each grid
function is double valued at  the patch interface coordinate since the
SBP derivative operators  are one sided at the  boundaries.  To ensure
consistency without compromising (linear)  stability, we make use of a
penalty  method as  described  above.  Constraint-preserving  boundary
conditions  require the  calculation  of derivatives  of certain  grid
functions at the outer boundary, which we also obtain by using the SBP
derivative operators.

        In a black hole setting,  the computational domain next to the
excision boundary tends to quickly amplify high frequency noise, which
can  not be  represented  accurately  on the  discrete  grid. This  is
especially true  for high  order accurate derivative  operators. Thus,
high  order  simulations of  black  holes  need  a certain  amount  of
numerical dissipation to be  stable. This dissipation is here provided
by    the   SBP   dissipation    operators   constructed    in   Ref.\
\cite{Diener05b}. The  free parameters of these  operators, namely the
coefficient of the dissipation and the extent of the transition region
(for non-diagonal operators), are found by numerical experiment.

\section{Results}
\label{sec:results}

The  numerical experiments  presented in  this section  are set  up to
systematically test  the performance of the new  techniques in several
situations of increasing  difficulty. We start with a  series of tests
evolving   a  Schwarzschild   black   hole  in   Painlev\'e-Gullstrand
coordinates with either a single  patch or two patches matched via the
penalty method, and compare the  performance of all SBP operators with
the   second    order   finite-differencing   method    presented   in
\cite{Calabrese02d}. Next, to test  more dynamical situations, a gauge
or scalar  field signal is injected in  a constraint-preserving manner
through the outer boundary and  accreted onto the black hole. A robust
stability test is then performed with noise on the incoming gauge mode
$u_3$, and,  with Cauchy--perturbative  matching, on the  scalar field
mode $u_8$.   Finally, a series of high-precision  tests involving all
techniques  are presented,  in which  a black  hole accretes  a scalar
field injected  through the outer boundary of  the perturbative patch.
These simulations also  include a test of the  long-term stability and
accuracy after accretion and ring-down.

\subsection{Schwarzschild black hole in Painlev\'e-Gullstrand coordinates}
\label{sec:bh_pg}

        In our  first series of  tests, a Schwarzschild black  hole is
evolved with high-order  accurate SBP operators, constraint-preserving
boundary  conditions and  excision.  Cauchy--perturbative  matching is
not used in these tests.  To fix the coordinate system, we make use of
the      horizon-penetrating     Painlev\'e-Gullstrand     coordinates
\cite{Gullstrand:1922,  Martel:2000rn},  and  we  fix  the  coordinate
functions  $\tilde{\alpha}$ and  $\beta$  of the  previous section  to
their exact values.

        For all tests,  the inner boundary is located  well inside the
event  horizon (more  precisely, it  is located  at $r_e=1  M$), which
implies that all modes  are outflow. Therefore, no boundary conditions
may be applied  at the excision boundary. The  exact boundary location
is not crucial as long as  it is inside the apparent horizon, but this
choice  facilitates  comparison  with \cite{Calabrese02d}.   Also,  in
dynamical   situations  the   apparent  horizon   location   may  move
significantly on the coordinate grid, and to ensure outflow conditions
at  the inner  boundary some  penetration into  the black  hole  is of
advantage.   To match  the setup  of \cite{Calabrese02d},  we  set the
outer boundary to $r=10 M$.  To ensure well-posedness of the continuum
problem, boundary  conditions should be applied to  the incoming modes
$u_1$, $u_2$, $u_3$, $u_4$, and  $u_8$. However, three of these modes,
namely  $u_1$,  $u_2$,  and  $u_4$,   can  be  fixed  by  the  use  of
constraint-preserving  boundary conditions,  as  discussed in  Section
\ref{sec:equations},  which leaves the  freely specifyable  gauge mode
$u_3$ and the scalar field mode $u_8$. Since in these initial tests we
are only interested in obtaining a stationary black hole solution, the
initial  scalar  field  is  set   to  zero,  and  the  (scalar  field)
characteristic mode $u_8$ is penalized  to zero as well.  The incoming
gauge mode $u_3$ is penalized to the exact solution.

        An  error function $\delta  M$ can  be defined  by use  of the
Misner-Sharp mass function \cite{Misner73}
\begin{equation}
M(r)  := \frac{r  g_T}{2} \left[  1 +  \frac{r^2}{g_T} \left(  K_T^2 -
\frac{f_{rT}^2}{g_{rr}} \right) \right],
\end{equation}
where then, if  the black hole mass is denoted by  $M$, $\delta M(r) =
(M(r)-M)/M$.   Since the same  error measure  and continuum  system is
used  in \cite{Calabrese02d},  we can  compare the  different discrete
approaches directly.

\subsubsection{One grid patch}

\begin{figure}
\includegraphics[width=\columnwidth]{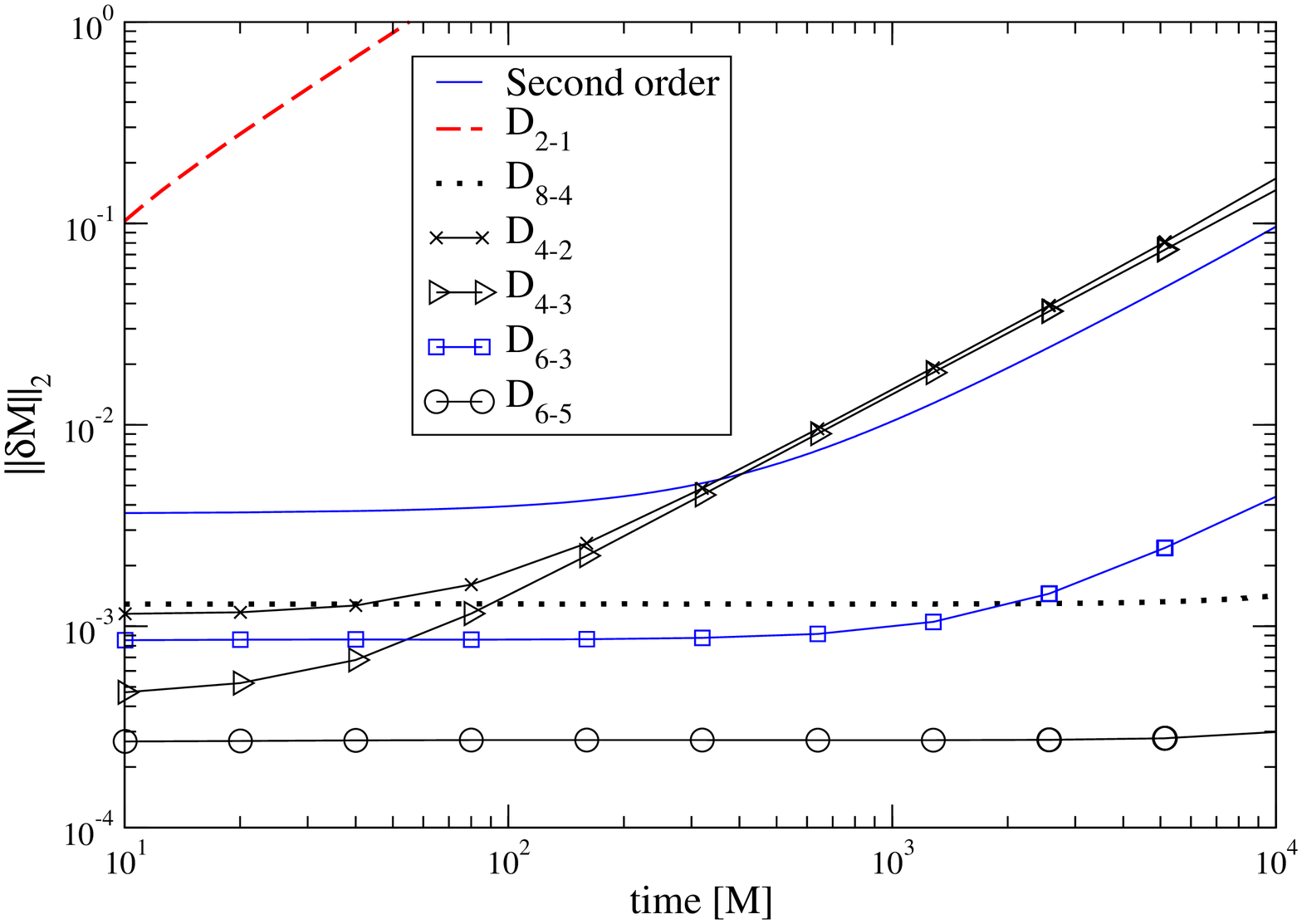}
\includegraphics[width=\columnwidth]{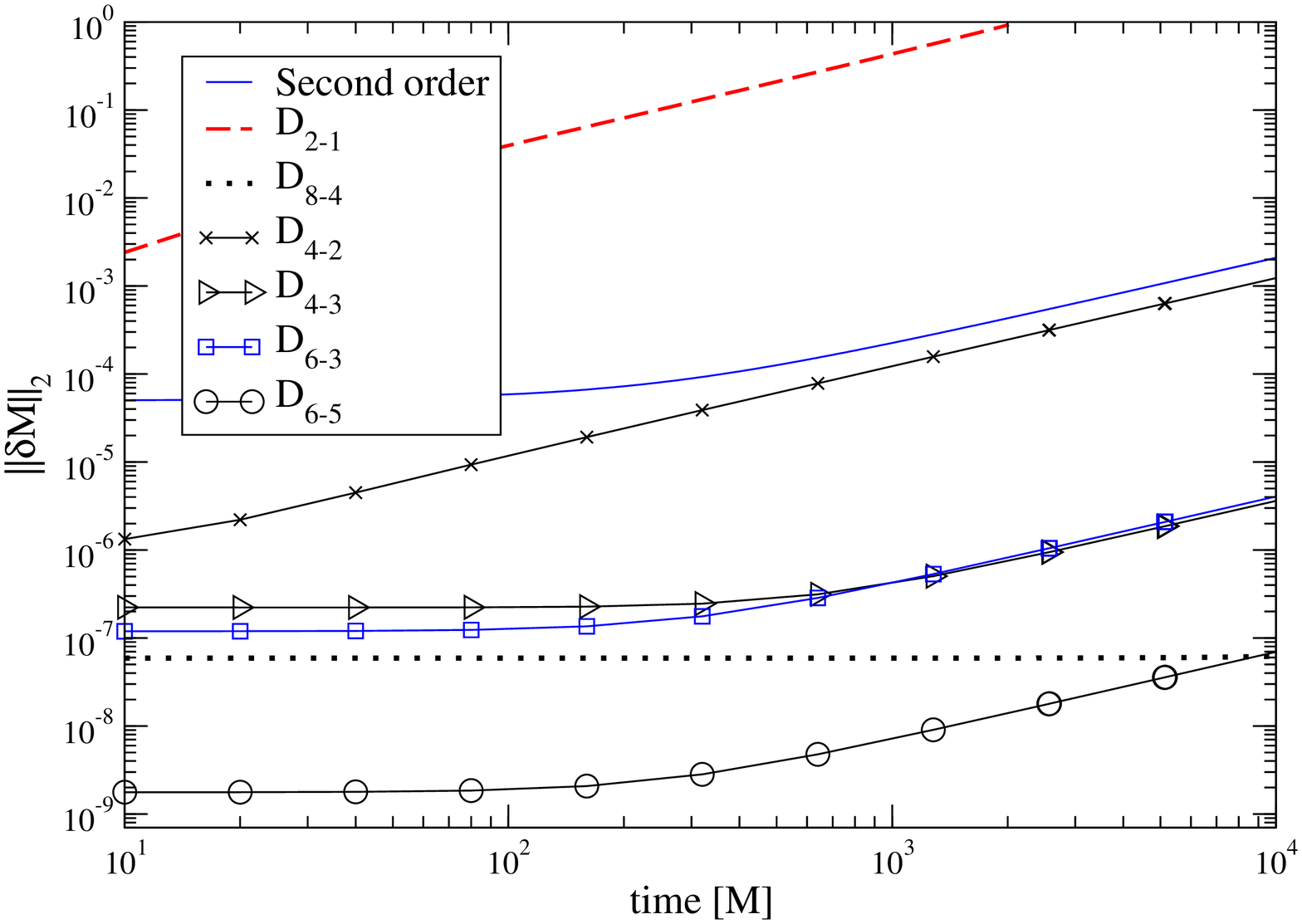}
\caption{Time evolution of the relative error in the Misner-Sharp mass
function    when   evolving    a   Schwarzschild    black    hole   in
Painlev\'e-Gullstrand coordinates  with one grid  patch, for different
discrete  methods.  Two  resolutions are  displayed,  corresponding to
$\Delta  r  =  M/8$  (upper  panel)  and  $\Delta  r  =  M/64$  (lower
panel). The result from the method presented ref.\ \cite{Calabrese02d}
is denoted  by ``second order'', while  new results are  marked by the
SBP  derivative  and   dissipation  operators  used.   The  high-order
operators $D_{6-5}$ and $D_{8-4}$ display superior performance already
at the lowest resolution.}
\label{fig:bh_pg} 
\end{figure}

The  computational  domain  $r  \in  [1,10]$  is  represented  by  one
coordinate  patch,  which  is  exactly  the same  setup  as  in  ref.\
\cite{Calabrese02d}.  In Figure~\ref{fig:bh_pg}  we compare for coarse
and high  resolutions, $\Delta r  = M/8,M/64$, the performance  of the
methods used in ref.\ \cite{Calabrese02d}, namely second order spatial
derivatives with fourth order  Kreiss-Oliger dissipation (which is set
to zero  near the boundaries) and  a third order  extrapolation at the
boundaries,  with   the  SBP  derivative   and  dissipation  operators
$D_{2-1}$,    $D_{4-2}$,   $D_{4-3}$,    $D_{6-3}$,    $D_{6-5}$   and
$D_{8-4}$. The  figure shows  the evolution of  the $L_2$ norm  of the
Misner-Sharp mass error  over an evolution time of  $10,000 M$. In all
cases displayed there is a linear growth in the error after some time.
This   is  an  artefact   of  the   discrete  representation   of  the
constraint-preserving  boundary conditions.   We  have also  performed
tests with  maximally dissipative  boundary conditions: these  yield a
discrete equilibrium after some time, and thus allow for evolutions of
unlimited  time.  However,  since  these boundary  conditions are  not
correct for  most systems of practical  interest, we only  make use of
this result  to point out  the source of  the linear growth  of errors
observed, which converges away with increasing resolution.

As  soon as  the  error gets  close  to $1$,  the  code encounters  an
instability, which,  in this case,  is associated with a  migration of
the  excision boundary  outside of  the black  hole,  and consequently
ill-posedness of the continuum problem.  While this migration could be
theoretically avoided by  choosing horizon-fixing dynamical coordinate
conditions, a solution  with this magnitude of error  is, in any case,
not of practical use.

        In the present numerical code, the SBP operators are also used
as  one-sided derivatives  for  determining the  constraint-preserving
boundary conditions, which suggests that the operator $D_{2-1}$, which
is only first order at  the boundaries, will yield less accurate outer
boundary    conditions    than    the    third   order    method    in
\cite{Calabrese02d}. Figure  \ref{fig:bh_pg} clearly demonstrates this
fact. However,  the operators  $D_{6-3}$, $D_{6-5}$ and  $D_{8-4}$ are
significantly  more  accurate  than  the results  presented  in  ref.\
\cite{Calabrese02d},  and  already  so  at  the  coarsest  resolution.
Furthermore, at $\Delta r = M/64$ the SBP operator $D_{6-5}$ induces a
solution error  of less than $10^{-7}$  (that is, {\em  four orders of
magnitude smaller than the  corresponding errors when using the second
order method of \cite{Calabrese02d}  with the same resolution}) within
$10,000  M$, which  appears sufficiently  accurate for  many practical
purposes.

\begin{figure}
\includegraphics[width=\columnwidth]{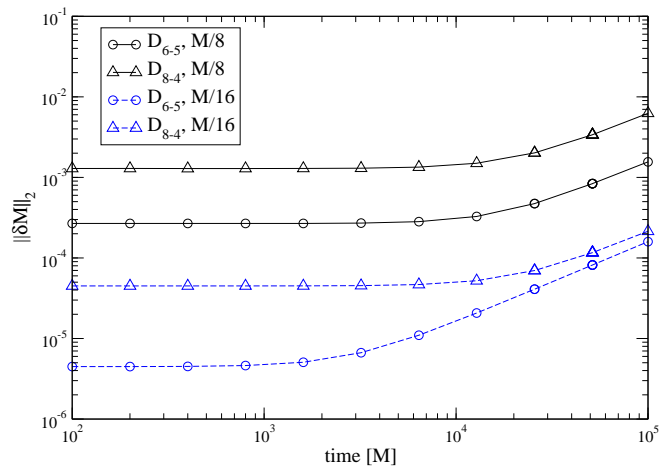}
\caption{Evolution of a Schwarzschild  black hole for $100,000 M$. The
axes show  the quantities described in  Figure~\ref{fig:bh_pg}.  It is
clear that even  with low resolutions of $\Delta r  = M/8$ and $M/16$,
the operators  $D_{6-5}$ and  $D_{8-4}$ are able  to evolve  the black
hole in a stable manner for a significant time.}
\label{fig:bh_pg_100000} 
\end{figure}

        The long-term evolution of a Schwarzschild black hole with the
operators     $D_{6-5}$    and     $D_{8-4}$    is     displayed    in
Figure~\ref{fig:bh_pg_100000}. The  linear growth of  errors dominates
the  solution  at  late  times,  but since  this  error  significantly
decreases with  resolution, long evolution times can  be obtained even
for moderate  radial grid spacings.  This is  naturally an interesting
feature  for simulations with  three-dimensional spatial  grids, where
computational resources are still a viable concern.

\subsubsection{Two grid patches}
\label{sec:bh_pg_two_patches}
        
        As  dicussed   in  the  introduction,  the   use  of  multiple
coordinate  patches has  advantages  when modelling  black holes.   To
implement a stable interface boundary condition, the penalty method is
used  to  ensure linear  stability.   Here  we  first investigate  the
performance  of the SBP  operators coupled  to an  inter-patch penalty
boundary  method by  evolving a  black hole  spacetime covered  by two
non-intersecting spherical shells, the first one  from $r = 1 M$ to $r
= 5.5 M$, and the second one from $r  = 5.5 M$ to $r = 10 M$. In order
to provide an  intermediate test towards the CPM tests  below, we do a
non-linear matching, communicating  all characteristic modes (that is,
without  imposing   for  the  moment   constraint-preserving  boundary
conditions at the matching interfaces).

The  free  parameter  of   the  penalty  boundary  condition  $\delta$
introduced in  section~\ref{sec:sbp} is  set to the  dissipative value
$0$.   Only  the  operators  $D_{6-5}$  and  $D_{8-4}$  are  used  for
comparison to the results from the previous section.

\begin{figure}
\includegraphics[width=\columnwidth]{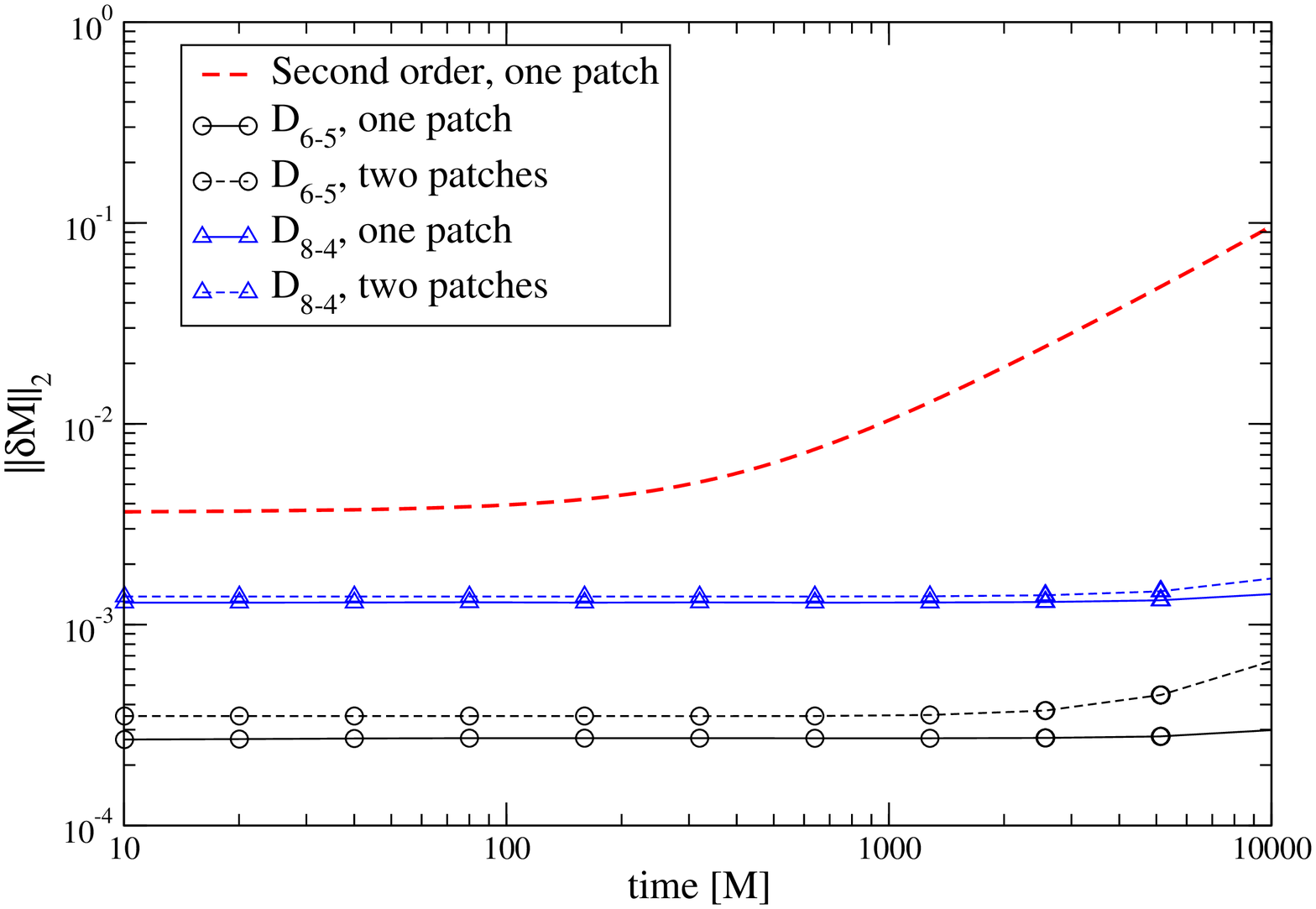}
\includegraphics[width=\columnwidth]{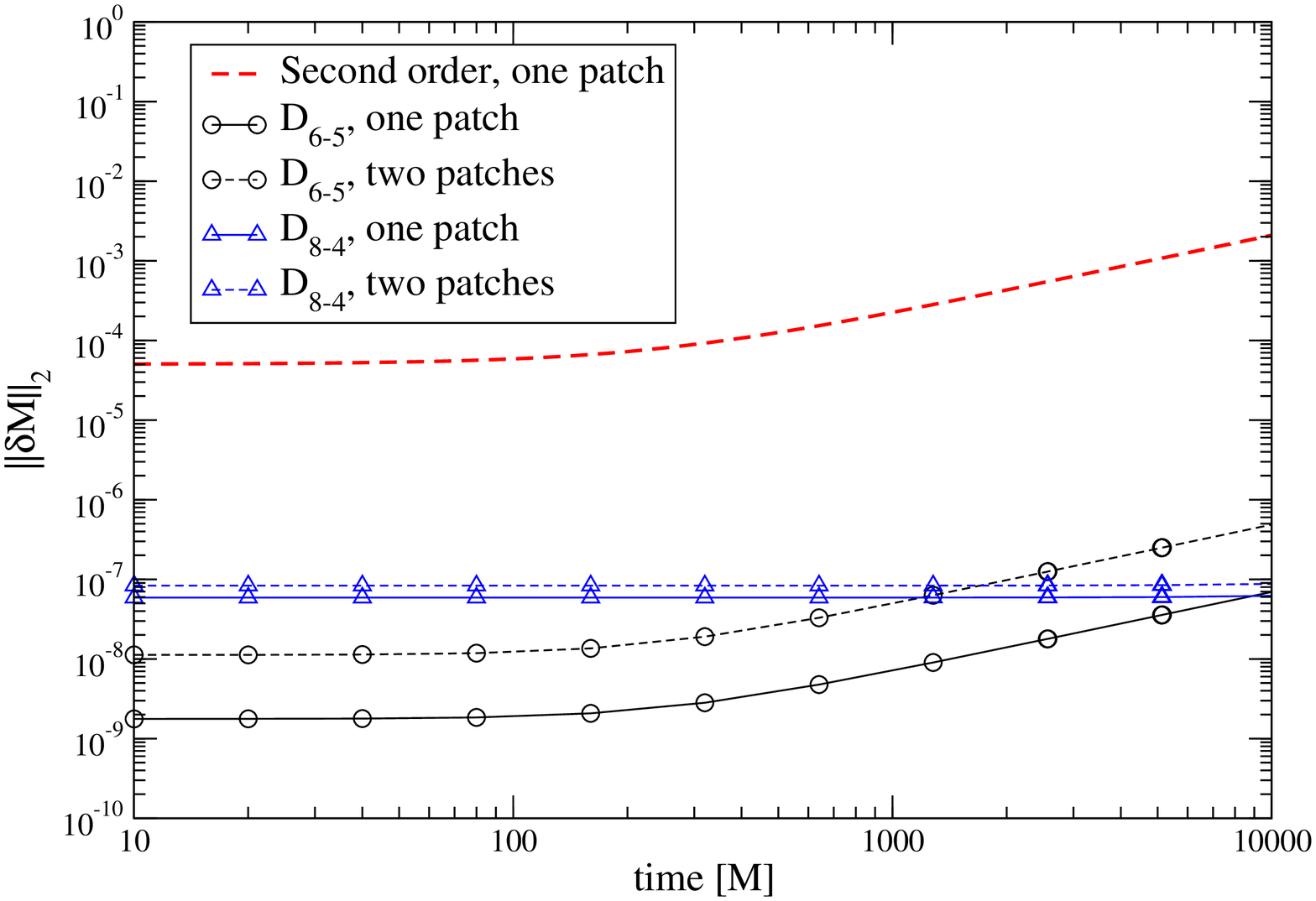}
\caption{Comparison  of  uni-patch  and  multi-patch evolutions  of  a
Schwarzschild  black hole  in Painlev\'e-Gullstrand  coordinates.  The
graphs denoted  by ``one patch''  and ``second order'' are  those from
Figure~\ref{fig:bh_pg},  while  the  corresponding  graphs  for  ``two
patches''  cover the  computational domain  with  two non-intersecting
spherical shells, the first one from $r = 1 M$ to $r = 5.5 M$, and the
second one from $r  = 5.5 M$ to $r = 10  M$. The one-sided derivatives
at the interface boundary introduce  a very small loss of accuracy. In
the upper  and lower panels the  resolution is $\Delta  r = M/8,M/64$,
respectively. For  the late time behaviour of  $D_{6-5}$ and $D_{8-4}$
please also cf. Figure~\ref{fig:bh_pg_100000}.}
\label{fig:bh_pg_two_patches} 
\end{figure}

\begin{figure}
\includegraphics[width=\columnwidth]{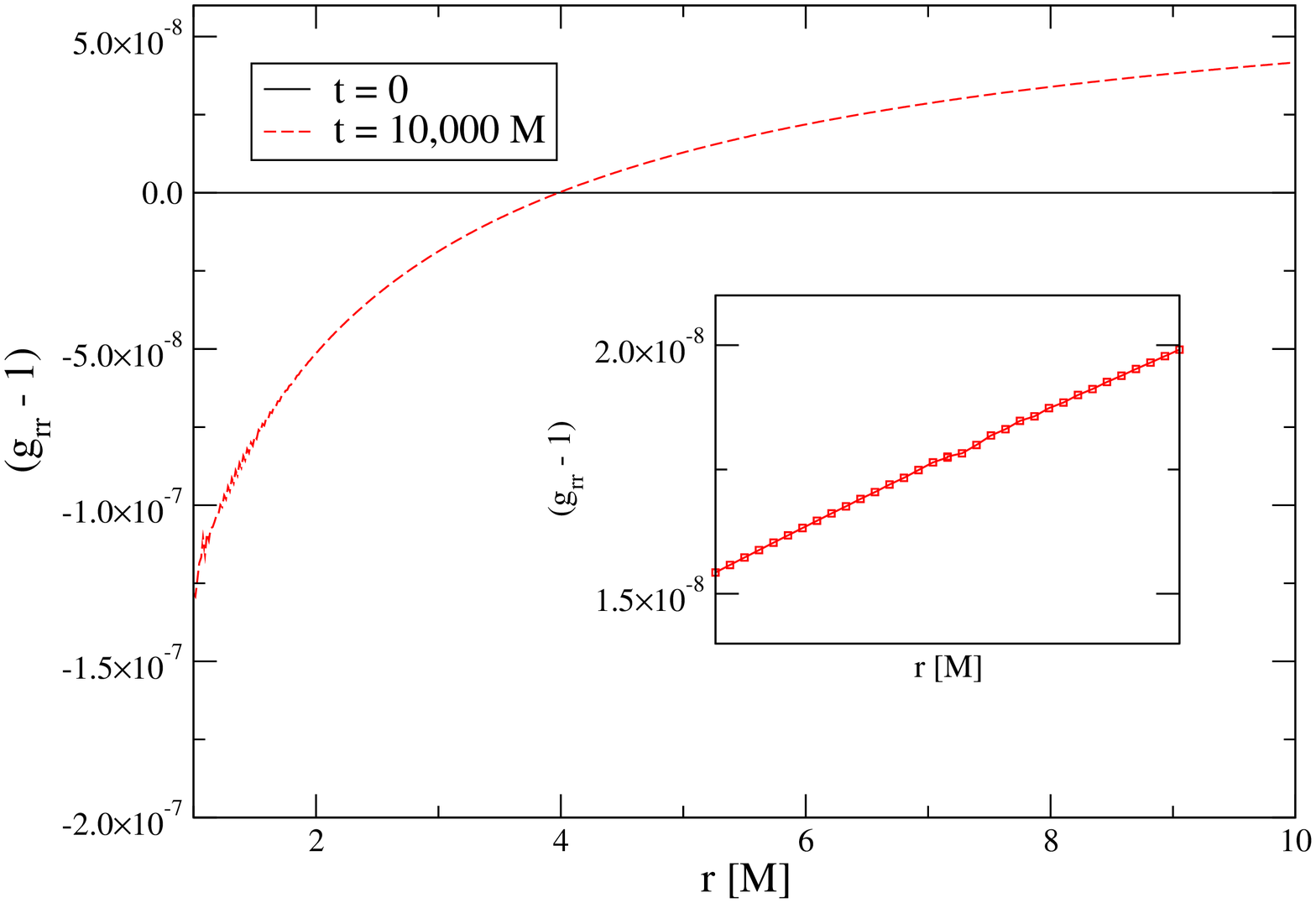}
\caption{Evolution  of metric function  $g_{rr}$ for  a black  hole in
Painlev\'e-Gullstrand coordinates,  with a  resolution of $\Delta  r =
M/64$, two grid patches with an interface at $r = 5.5 M$ and using the
SBP operator $D_{6-5}$. The two  graphs show the metric function at $t
= 0$ (where $g_{rr}(r)  = 1$) and at $t = 10,000  M$.  The inset shows
the region around the interface between the grid patches.}
\label{fig:bh_pg_two_patches_g_rr} 
\end{figure}

        In  Figure~\ref{fig:bh_pg_two_patches} the performance  of the
multi-patch  system is  compared  to the  uni-patch  results from  the
previous section. As expected, the use of one-sided derivatives at the
inter-patch boundary  reduces the  total level of  accuracy, but  in a
very  small  amount;  furthermore,  the  system is  still  stable  and
convergent. The time of the onset of the linear growth observed in all
evolutions  varies between  the grid  setups and  choices  of discrete
operator.  Figure  \ref{fig:bh_pg_two_patches_g_rr} shows the 3-metric
component $g_{rr}(r)$  at the times  $t = 0$  and $t = 10,000  M$. The
region around the inter-patch interface at $r = 5.5 M$ is shown in the
inset, which demonstrates that the penalty method introduces no strong
visible artifacts in this part  of the solution. This observation also
holds for the other solution functions.

\subsection{Gauge wave on a Schwarzschild background}
\label{sec:gauge_wave}

        The  next series of  tests focuses  on a  dynamical situation,
namely the  evolution of a Schwarzschild black  hole in non-stationary
coordinates.   For  this  purpose,  the  initial  data  is  set  to  a
Schwarzschild  black hole in  Painlev\'e-Gullstrand coordinates  as in
section (\ref{sec:bh_pg}), as  is the lapse and shift  function at all
times, but the incoming gauge mode  $u_3$ at the outer boundary is set
to a Gaussian pulse of the form

\begin{equation}
u_3(t) = u_3^{PG} (1 + A e^{-(t-t_0)^2/\tilde{\sigma}^2}).
\end{equation}

        Here, $u_3^{PG}$  is the exact gauge mode  from the stationary
solution. As  in ref.\ \cite{Calabrese02d},  we impose a  strong pulse
with $A  = 1$,  $t_0 =  5 M$ and  $\tilde{\sigma} =  2 M$.   Since the
solution  is now not  adapted to  the asymptotically  timelike Killing
field, the SBP operators and multi-patch techniques can be tested on a
solution  with wave propagation  without compromising  the use  of the
error measure  $||\delta M||_2$.  To facilitate  comparison with ref.\
\cite{Calabrese02d}, the  outer boundary is located  at $r =  30 M$ in
these tests.

\begin{figure}
\includegraphics[width=\columnwidth]{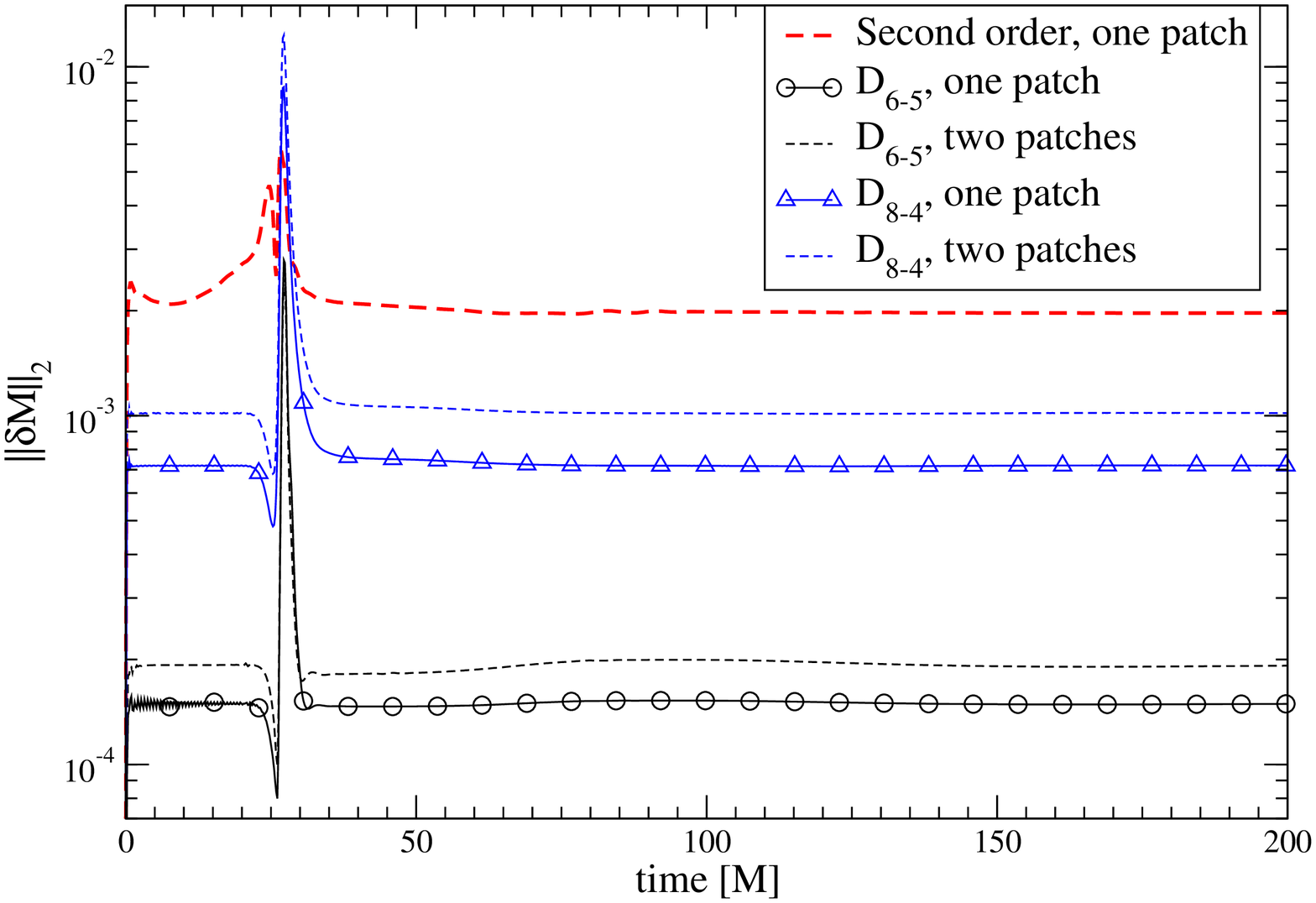}
\includegraphics[width=\columnwidth]{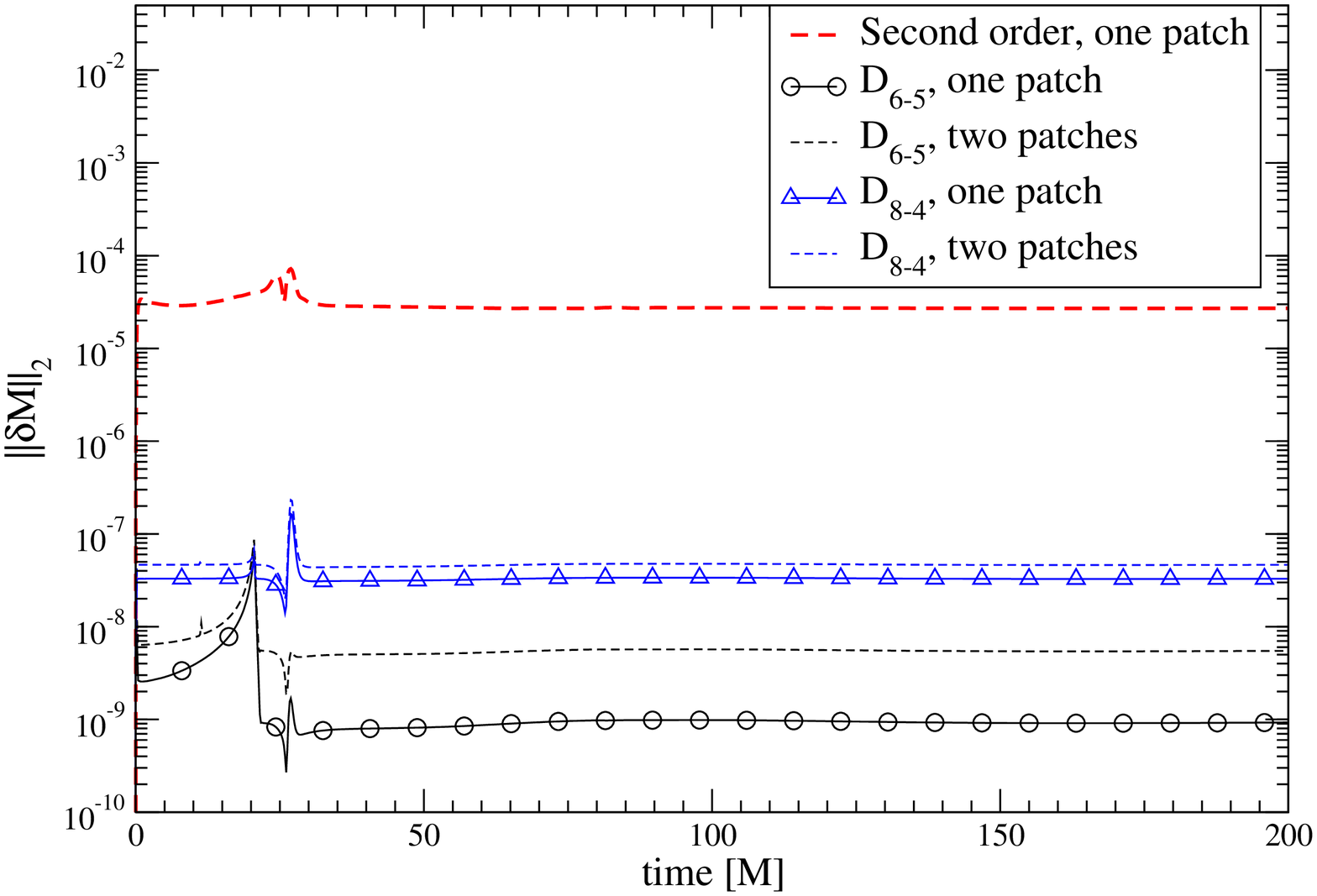}
\caption{Comparison of uni-patch and multi-patch evolutions of a gauge
wave travelling  on a Schwarzschild background. The  graphs denoted by
``second   order''   are   obtained   with  the   methods   in   ref.\
\cite{Calabrese02d}, while the  corresponding graphs for ``one patch''
and ``two patches'' cover the  computational domain with either one or
two non-intersecting spherical shells, the first one from $r = 1 M$ to
$r = 15.5 M$, and the second one from  $r = 15.5 M$ to $r = 30 M$. The
one-sided derivatives at the interface boundary introduce a small loss
of  accuracy, but  the system  is still  stable. The  upper  and lower
panels correspond to $\Delta r = M/8,M/64$, respectively.}
\label{fig:gauge_wave_two_patches} 
\end{figure}

        Figure \ref{fig:gauge_wave_two_patches} shows results from the
gauge pulse problem on a single  grid patch and two grid patches, here
with an  inter-patch boundary at  $r=15.5M$.  While in  the stationary
case  the inter-patch  boundary method  only  had to  deal with  small
numerically   introduced  differences  between   the  values   of  the
geometrical  quantities  at  the  interface, the  non-stationary  case
introduces a large  pulse travelling over the boundary,  and is thus a
much  more severe  test  for  accuracy and  stability  of the  penalty
method. The solution error is dominated by the ability of the discrete
method to represent the propagation  and accretion of the gauge pulse,
and by possible artefacts introduced by the inter-patch boundary.

         Judging   from  Figure~\ref{fig:gauge_wave_two_patches},  the
high-order operators are stable and significantly more accurate than a
second order method also in a dynamical situation, and even when using
multiple matched domains.

\subsection{Accretion of a scalar field pulse}
\label{sec:scalar_pulse}

        Since the  outer boundary has  two free incoming modes,  it is
possible to inject a scalar field  pulse in a way similar to the gauge
pulse  of section  (\ref{sec:gauge_wave}).  In  contrast to  the gauge
pulse, however, this system will result  in an increase of mass of the
black hole,  which also implies  that the Misner-Sharp mass  cannot be
used as a measure of the errors anymore. A possible choice for a gauge
field source with compact support is

\begin{displaymath}
u_8(t) = \left\{
\begin{array}{ll}
0  &  t<t_I  \\  \frac{A}{t_F^8}  (t-t_I)^4  (t-t_F)^4  \sin(\frac{\pi
t}{t_F}) & t \in [t_I,t_F] \\ 0 & t>t_F
\end{array}
\right.
\end{displaymath}

To  facilitate comparisons  with ref.\  \cite{Calabrese02d} we  use an
amplitude  $A =  7.2$, and  $t_I =  0 M$,  $t_F =  10 M$  and  set the
computational domain to be $r \in [1,50] M$.

\begin{figure}
\includegraphics[width=\columnwidth]{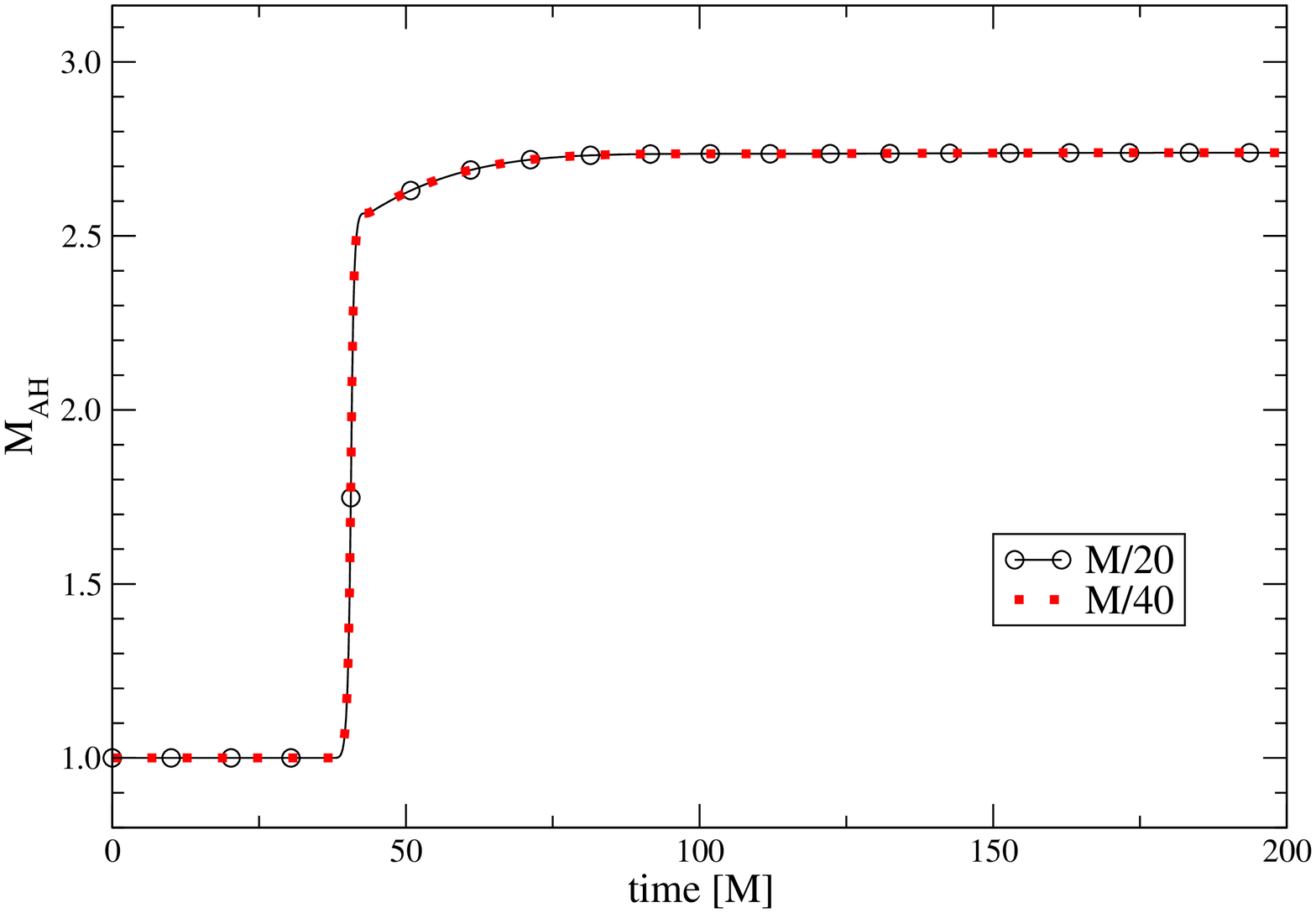}
\caption{Evolution of the apparent horizon mass for the accretion of a
strong scalar pulse to a Schwarzschild black hole. Shown are plots for
two resolutions,  $\Delta r = M/20$  and $\Delta r =  M/40$, using the
SBP operator  $D_{6-5}$. The large  scalar field amplitude leads  to a
significant increase in the black hole mass.}
\label{fig:ah_mass} 
\end{figure}

\begin{figure}
\includegraphics[width=\columnwidth]{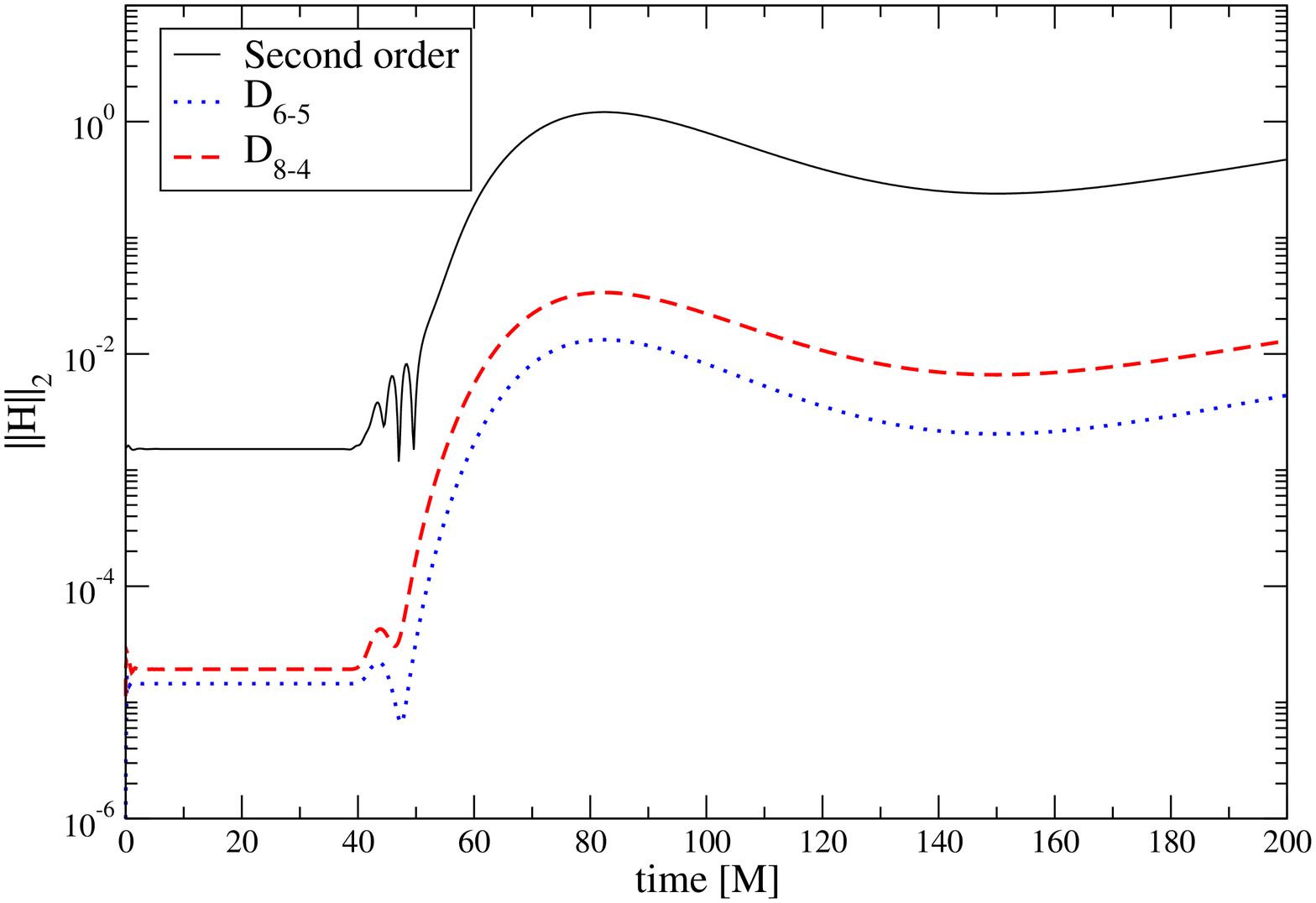}
\includegraphics[width=\columnwidth]{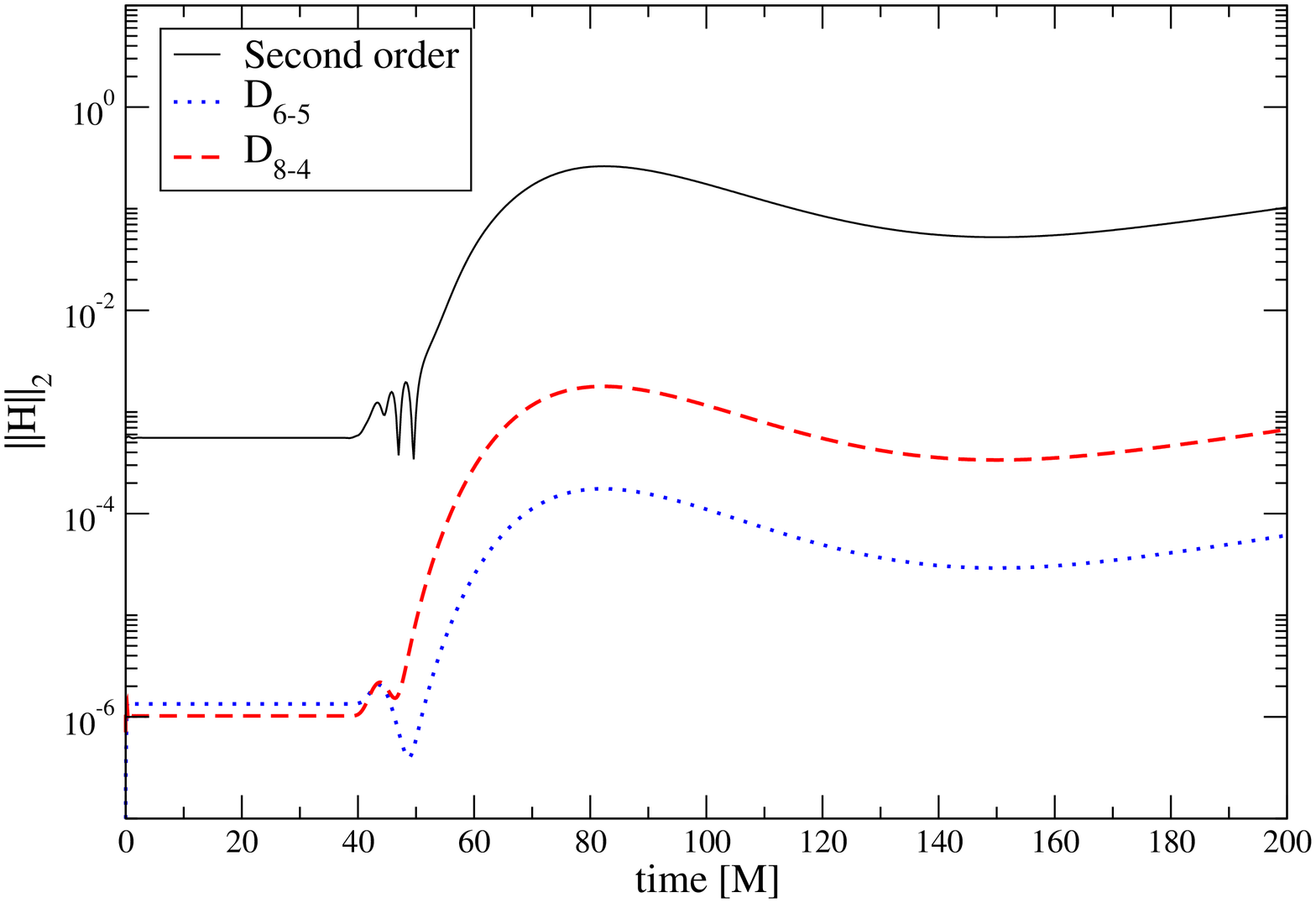}
\caption{$L_2$ norm  of the Hamiltonian  constraint over time  for the
accretion  of a  strong scalar  field pulse  to a  Schwarzschild black
hole, with resolutions $\Delta r = M/20,M/40$ (upper and lower panels,
respectively). The graph denoted  by ``second order'' is obtained with
the  method presented  in \cite{Calabrese02d},  and the  $D_{6-5}$ and
$D_{8-4}$ are obtained using the corresponding SBP operators.}
\label{fig:scalar_pulse_res_20} 
\end{figure}

\begin{figure}
\includegraphics[width=\columnwidth]{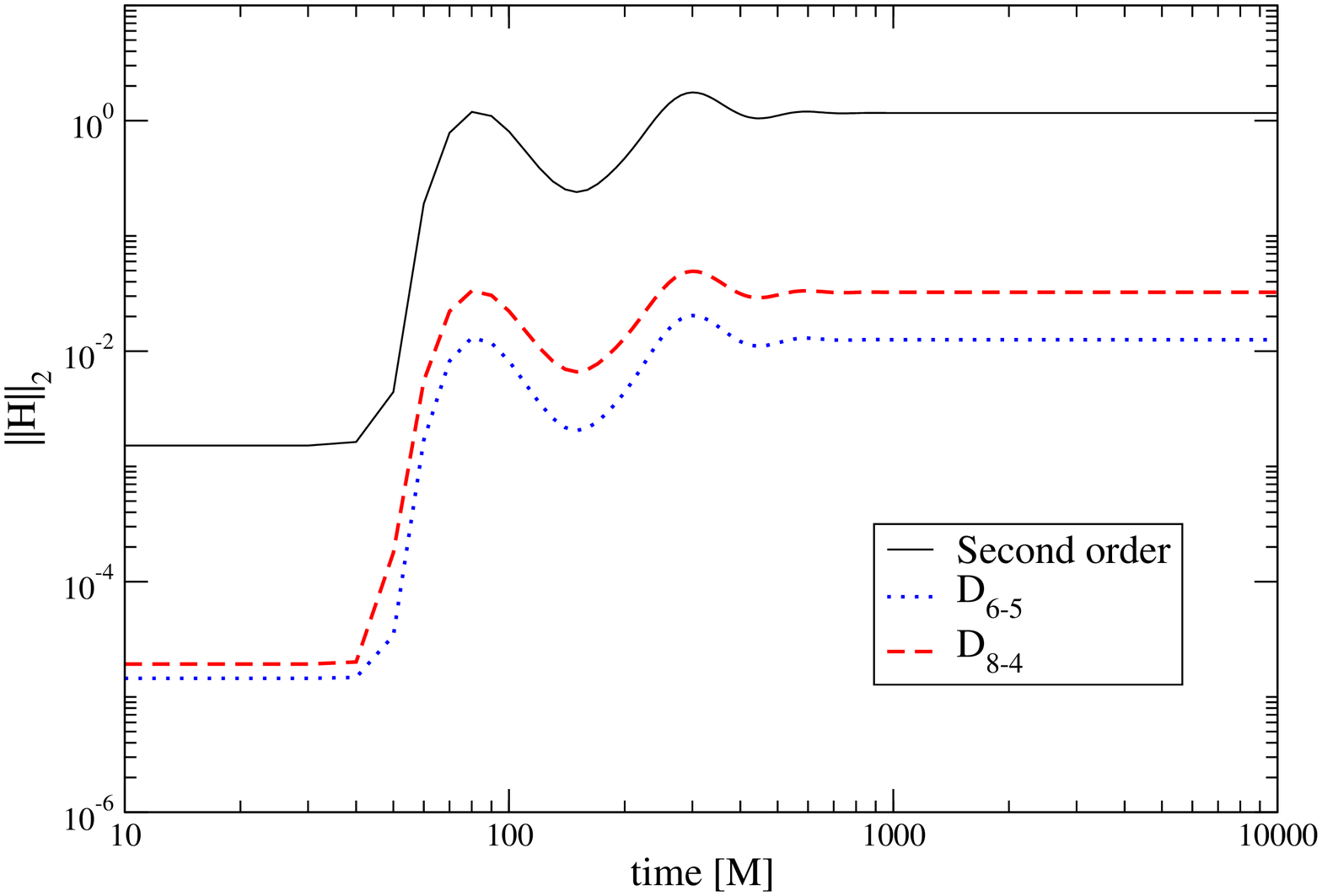}
\caption{As  Figure~\ref{fig:scalar_pulse_res_20},   but  evolved  for
10,000 M with $\Delta r = M/20$ to demonstrate the long-term behaviour
after accretion of the pulse.}
\label{fig:scalar_pulse_res_20_longterm} 
\end{figure}

        For resolutions $\Delta  r = M/20$ and $\Delta  r = M/40$, the
time    evolution   of    the   apparent    horizon   is    shown   in
Figure~\ref{fig:ah_mass}.   The scalar  pulse leads  to  a significant
increase in the black hole mass by a factor of $\approx 2.7$ after the
pulse is inside  the black hole. Larger amplitudes  are not obtainable
with the  simple gauge prescription used here,  but a horizon-freezing
gauge condition could improve on this result. As a replacement for the
Misner-Sharp error measure, we plot  the $L_2$ norm of the Hamiltonian
constraint over  time in Figure~\ref{fig:scalar_pulse_res_20}.   It is
apparent  that the  high-order  operators are  again  stable and  more
accurate than the second order operator.  The graphs indicate a growth
of the  constraint near $t  = 200 M$,  but a long-term  evolution with
$\Delta          r          =          M/20$         shown          in
Figure~\ref{fig:scalar_pulse_res_20_longterm}  demonstrates  that  the
system settles down to stability after the accretion.

\subsection{Robust stability test with gauge noise}
\label{sec:robust_stability_u3}

        The term \emph{robust stability test} \cite{Alcubierre:2003pc}
typically refers  to the discrete  stability of a numerical  system in
response to random perturbations.  In  this case, we will use the same
system as in  section (\ref{sec:bh_pg_two_patches}), but impose random
noise on the  incoming gauge mode $u_3$ with  a certain amplitude.  To
test the discrete stability of  the evolution system, we chose a large
range of amplitudes from  $10^{-4}$ to $0.3$.  Random perturbations of
the    latter   amplitude    is   significant    for    a   non-linear
system\footnote{Beyond  this  amplitude the  inner  boundary tends  to
become  partially inflow  by moving  the apparent  horizon  beyond the
computational  domain.   More sophisticated  gauge  or inner  boundary
condition could alleviate this, but  since we are interested here in a
proof of principle, a simple system is preferred.}.

\begin{figure}
\includegraphics[width=\columnwidth]{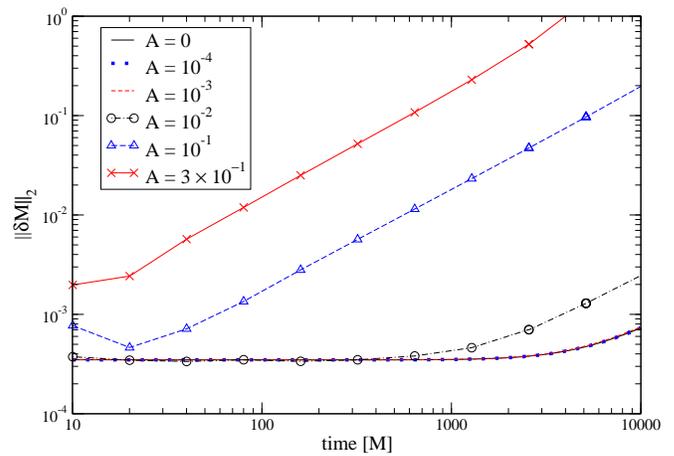}
\caption{Results of a robust stability test for different random noise
amplitudes.    The   system  is   a   Schwarzschild   black  hole   in
Painlev\'e-Gullstrand coordinates, and the computational domain $r \in
[1,10] M$ is covered by two patches with a boundary at $r = 5.5 M$ and
a resolution  of $M/8$.  Random  noise is superimposed on  the ingoing
gauge mode $u_3$,  with an amplitude denoted by  $A$.  The graphs show
the  mass  error with  time  for  different  random noise  amplitudes,
obtained with the SBP operator $D_{6-5}$.}
\label{fig:robust_stability_u3_amplitudes}
\end{figure}

\begin{figure}
\includegraphics[width=\columnwidth]{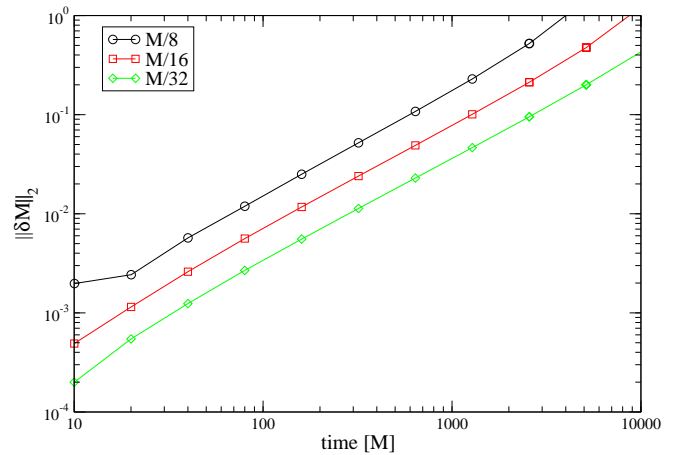}
\caption{Like Figure~\ref{fig:robust_stability_u3_amplitudes}, but for
the highest random noise amplitude $0.3$ and different resolutions.}
\label{fig:robust_stability_u3_resolutions}
\end{figure}

        For this  multi-patch test,  results in the  mass error  for a
resolution     $\Delta      r     =     M/8$      are     shown     in
Figure~\ref{fig:robust_stability_u3_amplitudes}.  It  is apparent that
strong random noise  induces a stronger growth in  the solution error.
However, this growth is still  linear. As in all black hole evolutions
in  section  (\ref{sec:bh_pg}),  the  system  encounters  a  numerical
instability as  the solution error approaches  $1$, but this  is not a
consequence of  the random noise,  but of the inner  boundary becoming
partially inflow due  to a coordinate motion of  the apparent horizon.
Also, with  increasing resolution, the  growth rate of the  error does
not            increase,             as            shown            in
Figure~\ref{fig:robust_stability_u3_resolutions}.   We  conclude  that
this high-order  evolution system is discretely  stable against strong
random perturbations.

\subsection{Cauchy--perturbative matching: robust stability test with scalar field noise}
\label{sec:robust_stability_u8}

        We   now    test   the   stability   of    the   system   with
Cauchy--perturbative  matching  against  random perturbations  in  the
scalar  field.   To  this  end,  the  computational  domain  is  again
subdivided  as  in  section (\ref{sec:robust_stability_u3}),  but  the
right patch evolves the  scalar field on a fixed Painlev\'e-Gullstrand
background as explained in  the introduction.  The interpatch boundary
is thus matching  the Cauchy patch to a perturbative  one, and we test
the stability  of the system against random  perturbations by imposing
random noise on  the incoming scalar field mode  on the outer boundary
of the perturbative patch.

\begin{figure}
\includegraphics[width=\columnwidth]{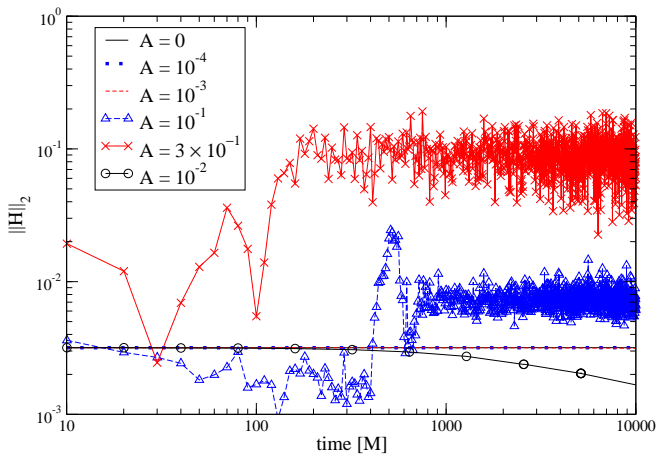}
\caption{Robust stability test with Cauchy--perturbative matching. The
system  is   a  dynamically   evolved  Schwarzschild  black   hole  in
Painlev\'e-Gullstrand coordinates matched  to a perturbative module at
$r = 5.5 M$ as described in the introduction.  Random noise is imposed
via the incoming scalar field  mode at the outer boundary.  Plotted is
the $L_2$ norm  of the Hamiltonian constraint over  time for different
noise  amplitudes.  All  evolutions  were done  with  a resolution  of
$\Delta r = M/8$ and the SBP operator $D_{6-5}$.}
\label{fig:robust_stability_u8_amplitudes}
\end{figure}

\begin{figure}
\includegraphics[width=\columnwidth]{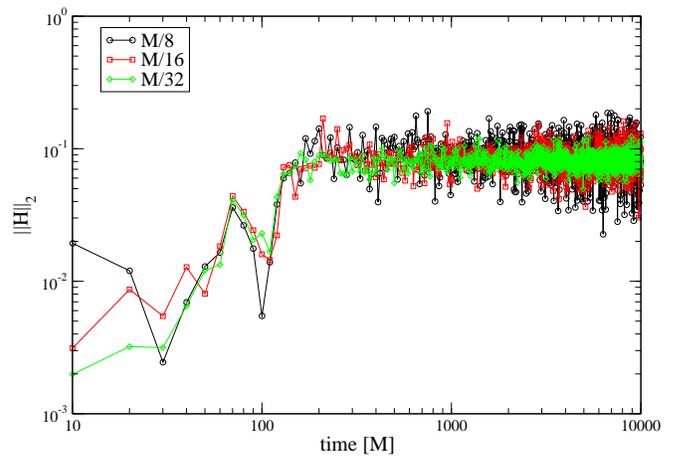}
\caption{Like Figure~\ref{fig:robust_stability_u8_amplitudes}, but for
the highest random noise amplitude and different resolutions.}
\label{fig:robust_stability_u8_resolutions}
\end{figure}

        Since the mass error is not available for a system accreting a
scalar field,  the $L_2$  norm of the  Hamiltonian constraint  is used
again    in    Figure~\ref{fig:robust_stability_u8_amplitudes}.     No
exponential  growth  can be  observed  in  the Hamiltonian  constraint
violation. The  same is  true when increasing  the resolutions,  as in
Figure~\ref{fig:robust_stability_u8_resolutions},  which also deserves
some additional comments: The robust stability test does not lead to a
converging sequence of solutions if  the random noise amplitude is not
diminished with resolution.  However, the purpose of these tests is to
excite  any unstable  high freqcency  modes present  in  the numerical
system.  The  absence of any  mode growing with  increasing resolution
shows that  the system with a  Cauchy--perturbative matching interface
is stable even  against strong random noise injected  into the system.
This  is a  promising result  for any  effort to  do three-dimensional
matching between  Cauchy modules and perturbative  ones using multiple
patches and high-order summation-by-parts operators.

\subsection{Cauchy--perturbative matching: Accretion of a ``gravitational wave''
and long-term evolution}
\label{sec:cpm_gw_accretion}

        Finally,  using the  massless Klein-Gordon  field as  a scalar
analogue of  gravitational waves in  spherical symmetry, we  model the
accretion of a gravitational wave packet across a Cauchy--perturbative
matching  boundary. This  test  is an  extension  of the  single-patch
scalar field accretion  of section (\ref{sec:scalar_pulse}), and makes
use  of all  ingredients  presented in  this  paper for  a stable  and
accurate evolution of black holes with Cauchy--perturbative matching.

\begin{figure}
\includegraphics[width=\columnwidth]{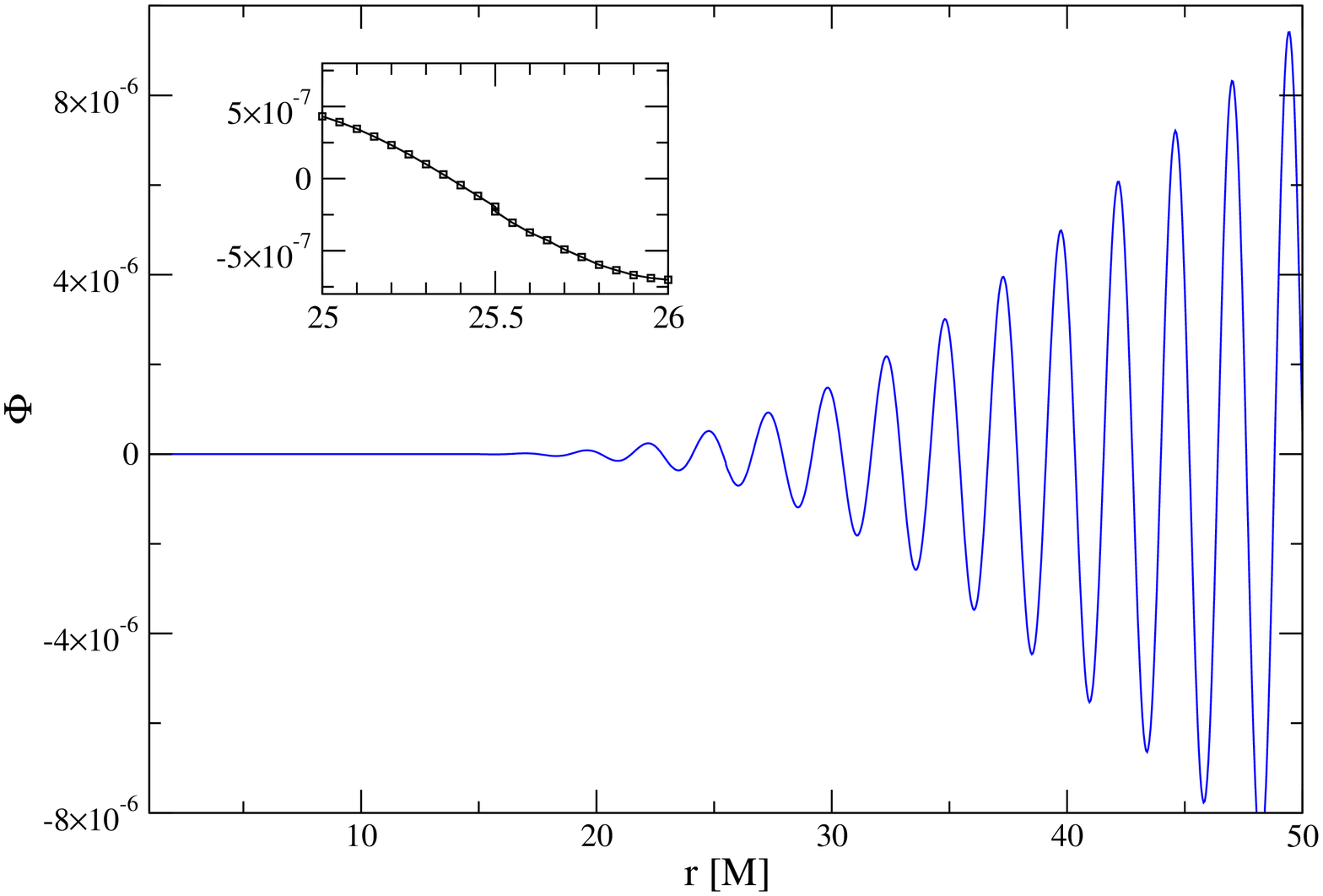}
\includegraphics[width=\columnwidth]{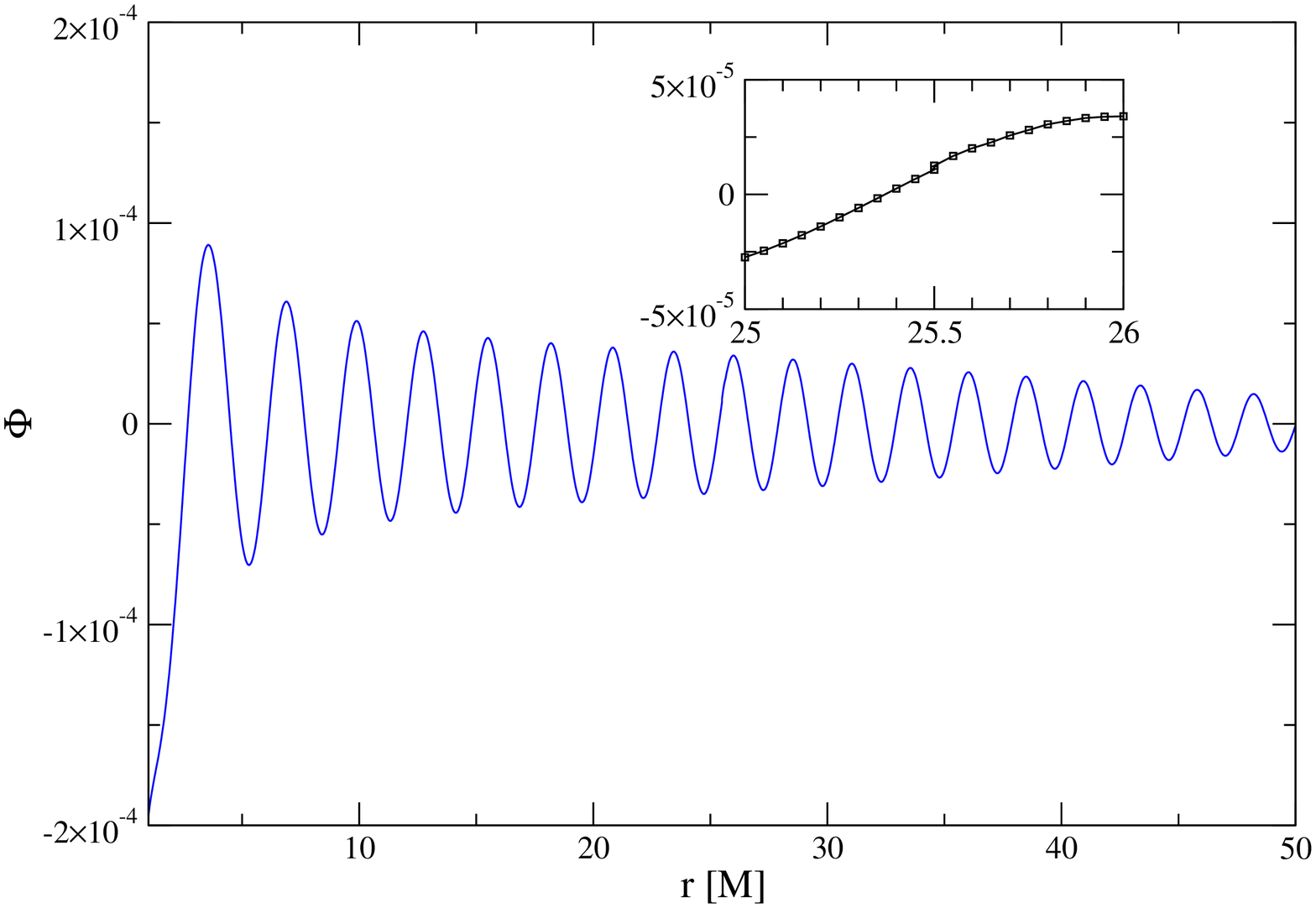}
\includegraphics[width=\columnwidth]{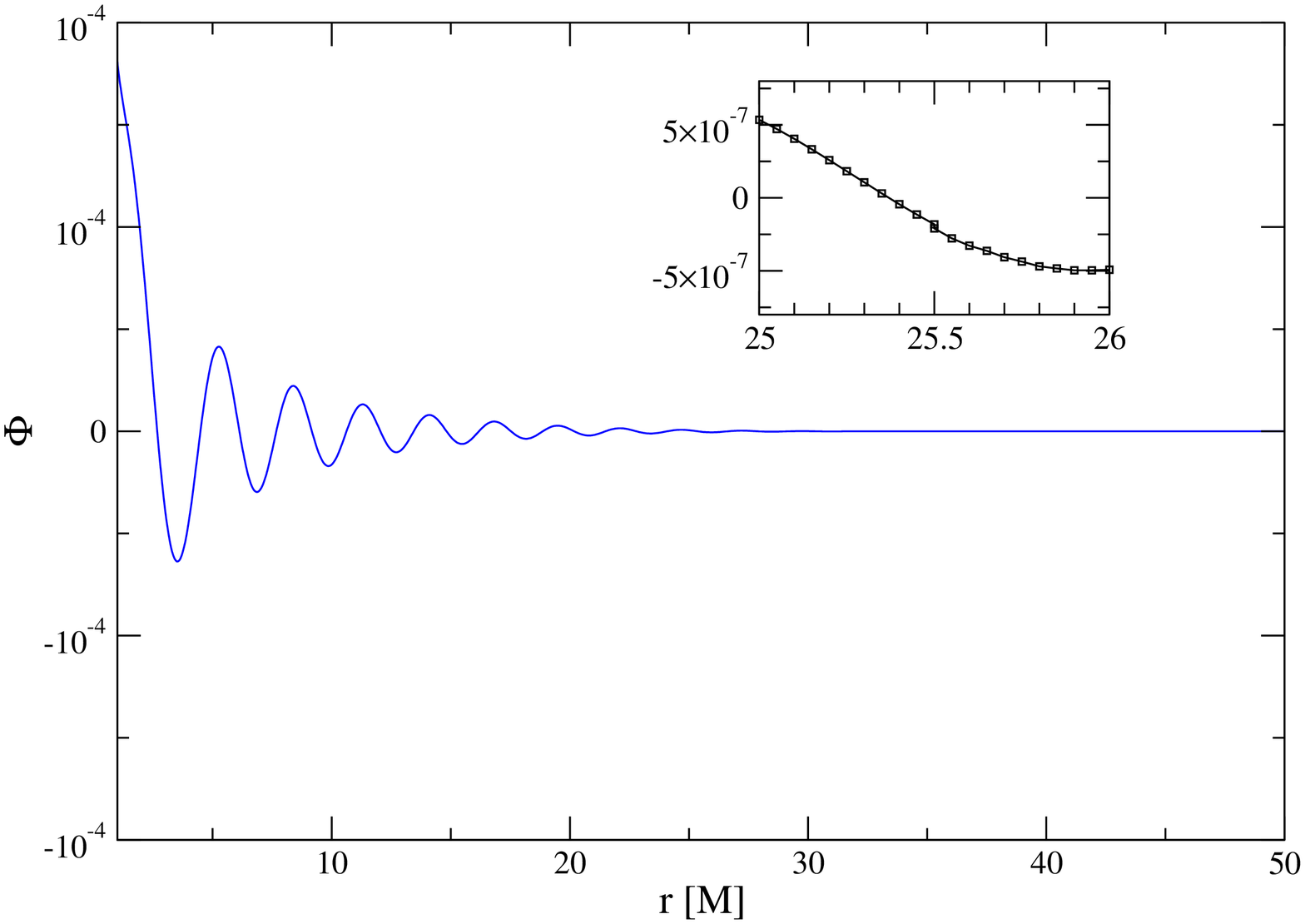}
\caption{Accretion    of    a   scalar    wave    packet   across    a
Cauchy--peturbative  matching  interface,   as  a  scalar  analog  for
gravitational  wave accretion  in three-dimensional  simulations.  The
packet consists of 50  waves injected from $t = 0$ to  $t = 100 M$, as
described in the text.  Here, the grid function $\Phi$ is plotted over
the radial coordinate  at $t = 30 M, 65M, 110M$  (from top to bottom),
for  the   resolution  $\Delta  r   =  M/20$  and  the   SBP  operator
$D_{6-5}$. The inset  shows the behaviour of the  grid function around
the  matching interface,  which is  at $r  = 25.5  M$. Note  that even
though the grid function is  in principle two-valued on the interface,
the penalties in conjunction with  high order operators only lead to a
very small mismatch.}
\label{fig:cpm_gw_accretion_phi}
\end{figure}

\begin{figure}
\includegraphics[width=\columnwidth]{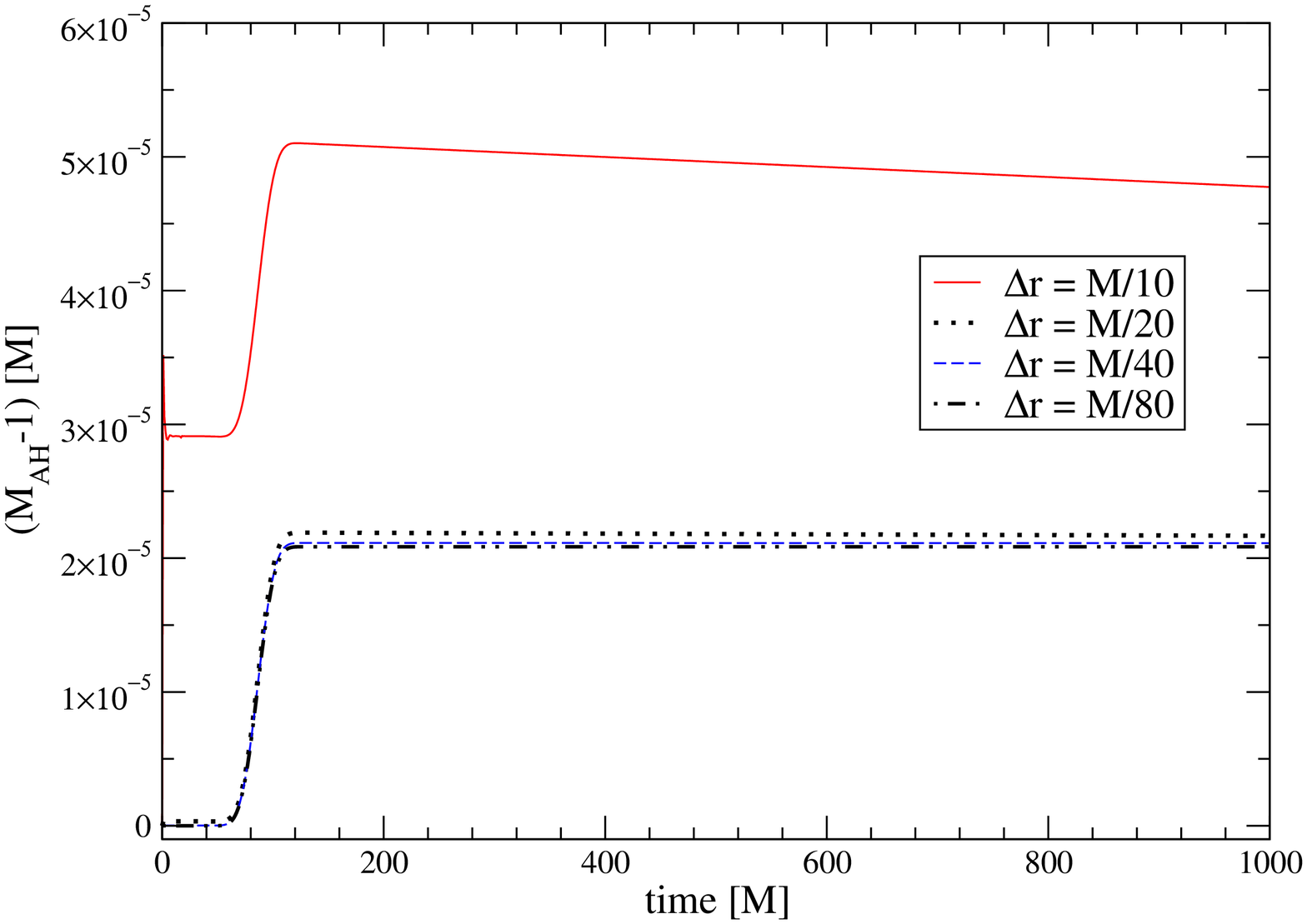}
\caption{Accretion    of    a   scalar    wave    packet   across    a
Cauchy--peturbative matching  interface. This plot  shows the apparent
horizon mass  over time for evolutions with  different resolutions and
the SBP operator $D_{6-5}$.}
\label{fig:cpm_gw_accretion_ah_mass}
\end{figure}

\begin{figure}
\includegraphics[width=\columnwidth]{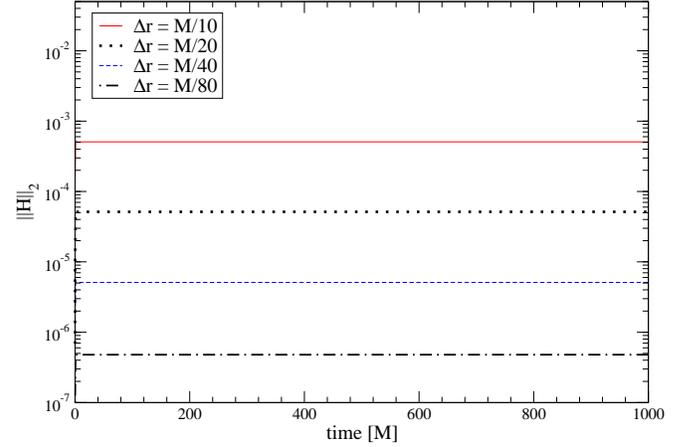}
\caption{Accretion    of    a   scalar    wave    packet   across    a
Cauchy--peturbative matching interface. This plot shows the $L_2$ norm
of the Hamiltonian constraint for different resolutions, using the SBP
operator  $D_{6-5}$.    {\em  The  non-linear   constraint  violations
introduced at the continuum by the matching are small enough that they
cannot be  detected in these very accurate  simulations.} Please note,
for   comparison  with  Figure~\ref{fig:scalar_pulse_res_20_longterm},
that the amplitude  of the Klein-Gordon signal is  smaller compared to
section (\ref{sec:scalar_pulse}).}
\label{fig:cpm_gw_accretion_ham}
\end{figure}

        Since Cauchy--perturbative  matching assumes the gravitational
wave to  be a  small perturbation  of a fixed  background in  the wave
zone,  the amplitude of  the wave  packet that  we inject  through the
outermost  boundary  is  chosen  to  be  $A  =  0.01$.   Similarly  to
section~\ref{sec:scalar_pulse}, we describe the packet by the function
\begin{displaymath}
u_8(t) = \left\{
\begin{array}{ll}
0 &  t<t_I \\ \frac{A}{t_F^8} (t-t_I)^4  (t-t_F)^4 \sin(\frac{\pi t}{n
t_F}) & t \in [t_I,t_F] \\ 0 & t>t_F
\end{array}
\right.
\end{displaymath}
where for the number  of half waves in the pulse we  set $n = 100$. We
inject  the packet  from $t_I  = 0$  to $t_F  = 100  M$. The  plots in
Figure~\ref{fig:cpm_gw_accretion_phi} display the the evolution of the
grid function  $\Phi$, and specifically the behaviour  of the function
around the  Cauchy--perturbative matching interface, which is  at $r =
25.5 M$. The corresponding increase  in apparent horizon mass is shown
in  Figure~\ref{fig:cpm_gw_accretion_ah_mass}.  The  evolution  of the
Hamiltonian constraint violation using  the SBP operator $D_{6-5}$ and
different          resolutions           is          shown          in
Figure~\ref{fig:cpm_gw_accretion_ham}.  It  is apparent that  with the
techniques used not  only is the discrete system  stable and accurate,
{\em  but   also  the  amount  of   non-linear  constraint  violations
introduced at  the continuum by the  Cauchy--perturbative matching are
very   small},  in  Figure~\ref{fig:cpm_gw_accretion_ham}   they  must
actually be smaller than $10^{-6}$.

\begin{figure}
\includegraphics[width=\columnwidth]{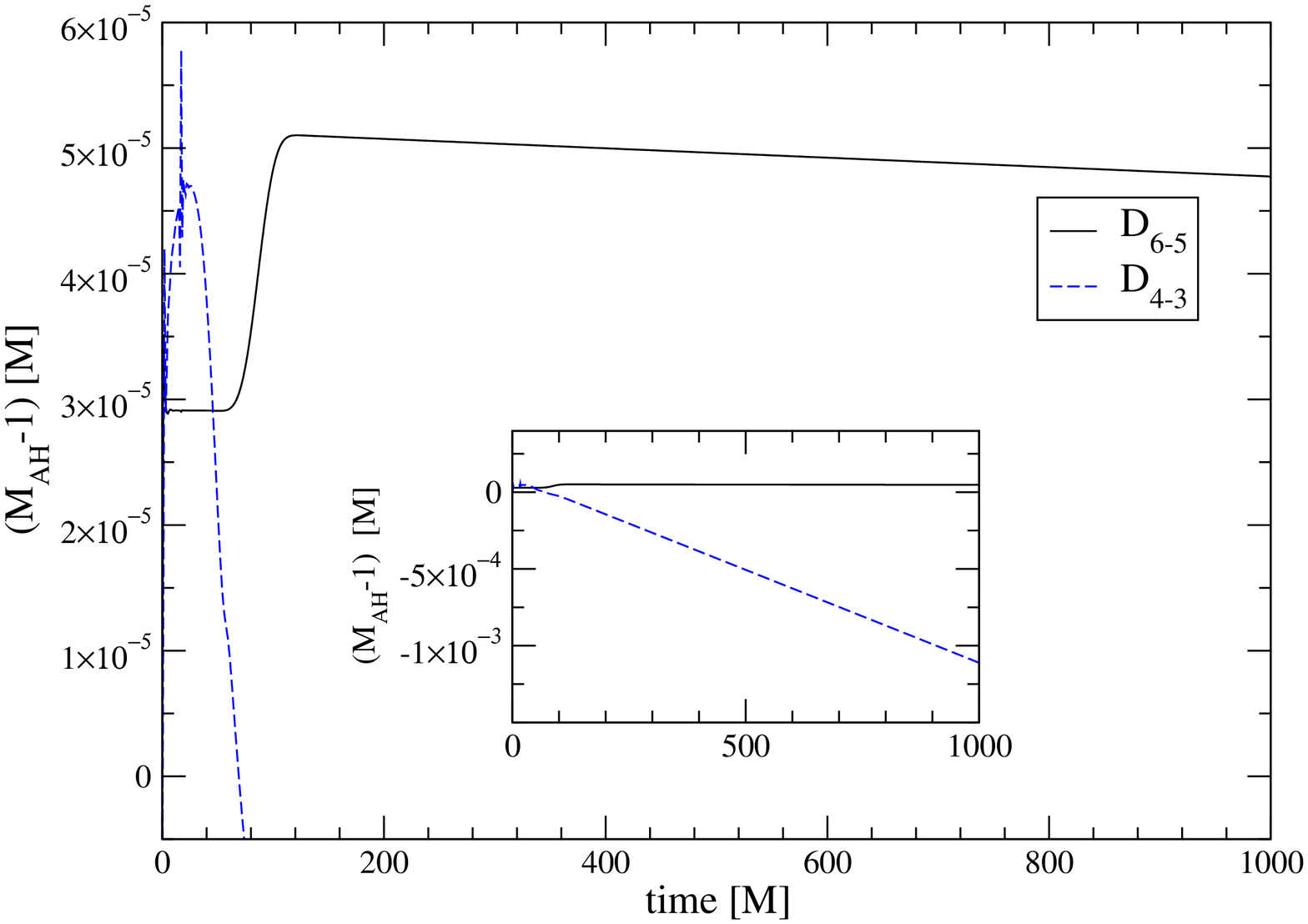}
\caption{Accretion    of    a   scalar    wave    packet   across    a
Cauchy--peturbative matching interface.   To demonstrate the advantage
of  using high-order  methods,  $(M_{AH}-1)$ is  shown for  evolutions
obtained with  the SBP operators $D_{4-3}$  $D_{6-5}$, with resolution
$\Delta r = M/10$. The loss of mass after accretion of the wave packet
with compact  support in  $t \in [0,100]  M$ is a  numerical artefact,
which  converges  away with  resolution.   The  inset  shows that  the
evolution obtained  with the operator  $D_{4-3}$ is not  unstable, but
only significantly less accurate.}
\label{fig:cpm_gw_accretion_ah_mass_res_10}
\end{figure}

\begin{figure}
\includegraphics[width=\columnwidth]{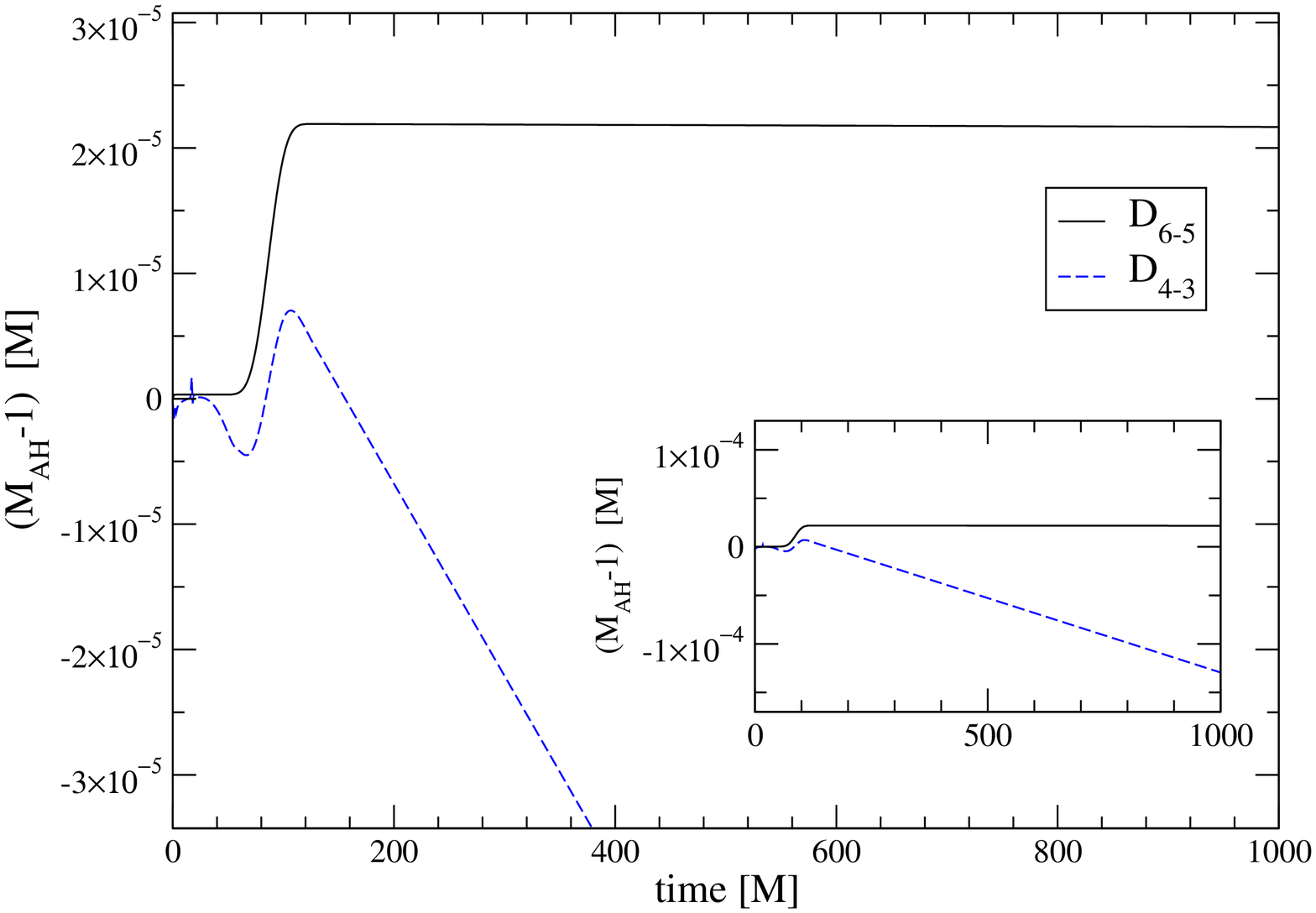}
\caption{Like  figure  \ref{fig:cpm_gw_accretion_ah_mass_res_10},  but
for a resolution of $\Delta r = M/20$.}
\label{fig:cpm_gw_accretion_ah_mass_res_20}
\end{figure}

\begin{figure}
\includegraphics[width=\columnwidth]{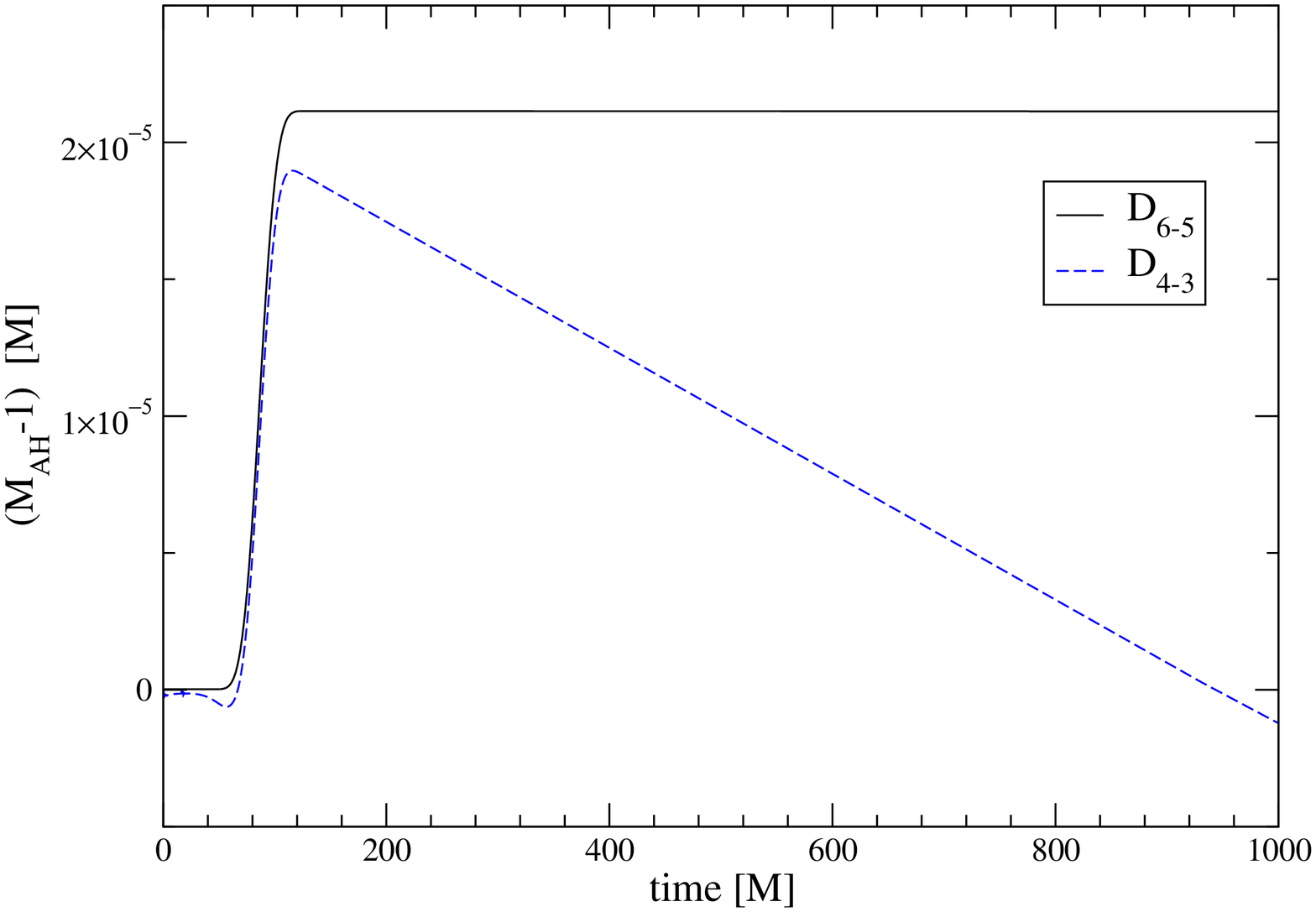}
\caption{Like   Figure~\ref{fig:cpm_gw_accretion_ah_mass_res_10},  but
for a resolution of $\Delta r = M/40$.}
\label{fig:cpm_gw_accretion_ah_mass_res_40}
\end{figure}

\begin{figure}
\includegraphics[width=\columnwidth]{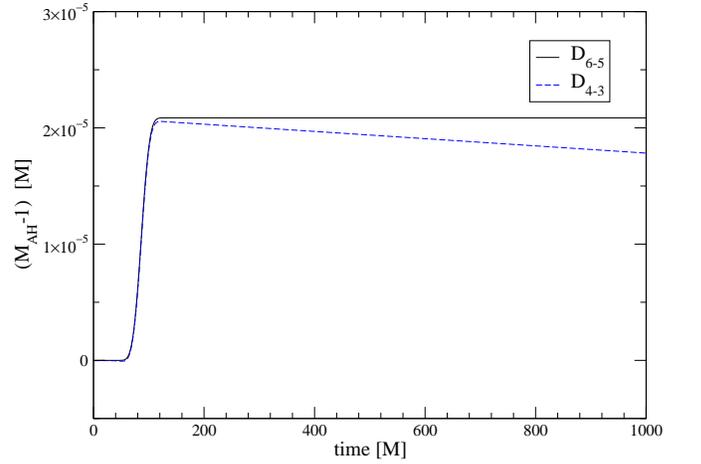}
\caption{Like   Figure~\ref{fig:cpm_gw_accretion_ah_mass_res_10},  but
for a resolution of $\Delta r = M/80$.}
\label{fig:cpm_gw_accretion_ah_mass_res_80}
\end{figure}

        The advantages of using  high-order methods is made evident in
Figures~\ref{fig:cpm_gw_accretion_ah_mass_res_10},
\ref{fig:cpm_gw_accretion_ah_mass_res_20},
\ref{fig:cpm_gw_accretion_ah_mass_res_40},                          and
\ref{fig:cpm_gw_accretion_ah_mass_res_80}.    In   these  plots,   the
performance of the SBP operator $D_{6-5}$, which is sixth order in the
interior and fifth order at the boundaries, is compared to that of the
operator $D_{4-3}$,  which is fourth  order in the interior  and third
order at the boundaries, for different choices of resolution. Although
both  operators  show  convergence,  for  a  mass  increase  of  about
$10^{-5}$, the  operator $D_{4-3}$ is unable to  reproduce the correct
behaviour  with   reasonable  grid  resolutions.    We  consider  this
specifically  important for  three-dimensional simulations,  where the
necessary resources scale with $n^4$ if $n$ denotes the number of grid
points  in each  direction.   Thus, for  all  simulations requiring  a
certain  amount of  precision, high-order  operators are  an essential
requirement.

\begin{figure}
\includegraphics[width=\columnwidth]{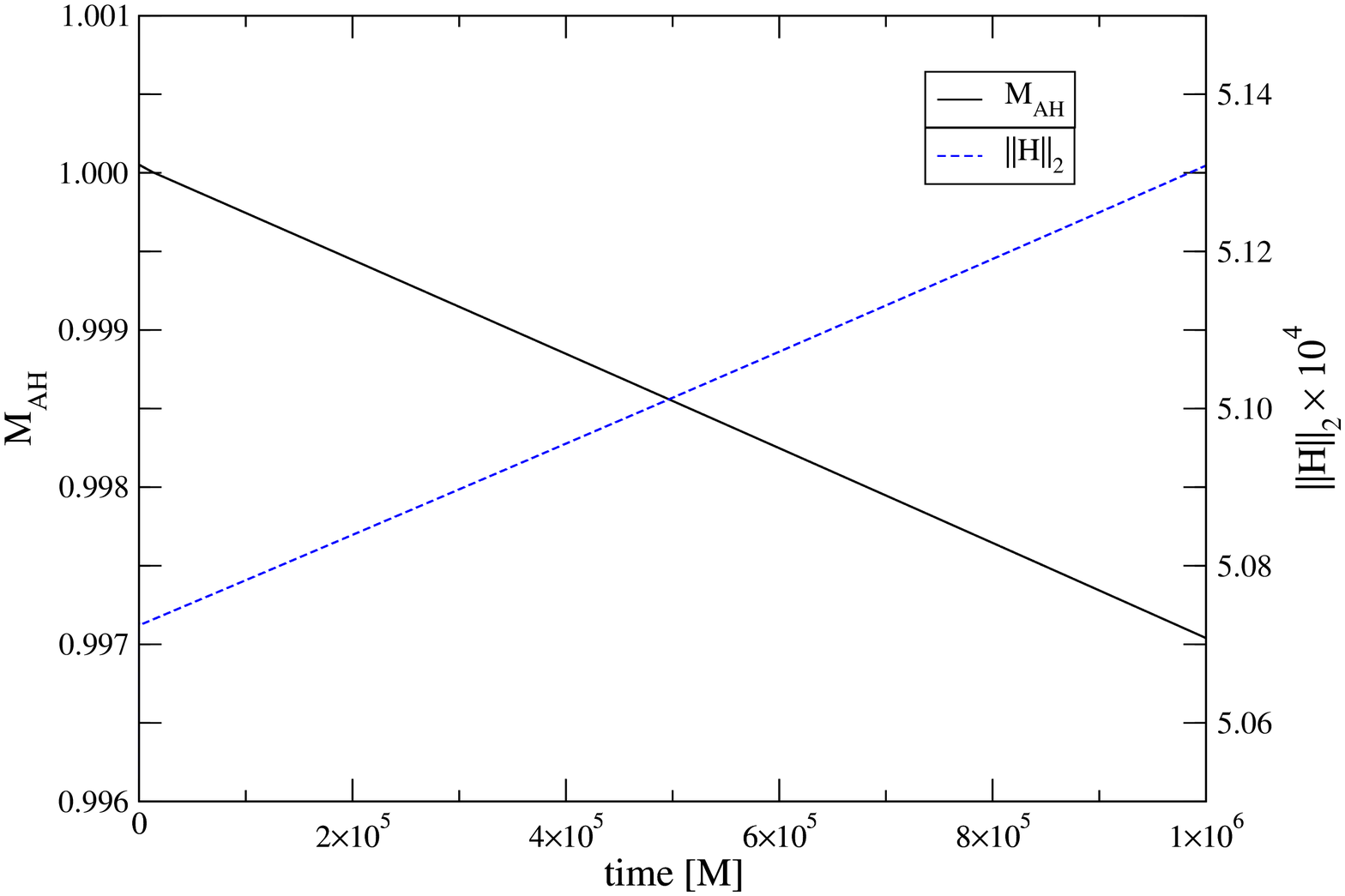}
\caption{Long-term  stable  evolution of  a  Schwarzschild black  hole
after  accretion of  a  scalar wave  packet with  Cauchy--perturbative
matching.  The  SBP operator  $D_{6-5}$ is used  with a  resolution of
$\Delta  r = M/10$.   Plotted are  the apparent  horizon mass  and the
Hamiltonian constraint over time.  The apparent horizon mass indicates
that  the discrete  evolution  introduces a  relative  error of  about
$0.3\%$ after $1,000,000 M$.}
\label{fig:cpm_gw_accretion_1e6}
\end{figure}

        The  long-term   evolution  of  a   Schwarzschild  black  hole
accreting a wave packet over a Cauchy--perturbative matching interface
and     settling     down    to     equilibrium     is    shown     in
Figure~\ref{fig:cpm_gw_accretion_1e6}.  The black  hole is evolved for
$1,000,000 M$ with the lowest resolution $\Delta r = M/10$ and the SBP
operator $D_{6-5}$. While an evolution  of this length might appear to
be of only  technical interest, we note that  modelling phenomena like
hypernovae  and  collapsars in  general  relativity  will require  the
stable evolution of  a black hole for at  least several seconds, which
is the lower end of  timescales associated with the collapsar model of
gamma-ray burst  engines \cite{Woosley01a}.  For a  stellar mass black
hole, $M  = M_\odot  \approx 5 \mu  s$, that  is $1 s  \approx 200,000
M_\odot$.

\section{Conclusions and outlook}
\label{sec:conclusions}

        To  obtain  long-term   evolutions  of  compact  astrophysical
systems in three spatial dimensions, advanced numerical techniques are
preferable  in that  they may  improve stability  and accuracy  of the
associated  discrete  model   system.   While  high  accuracy  enables
efficient use of the available computational resources, well-posedness
of the continuum model  and numerical stability are requirements which
can  not  be met  by  increasing  computational  power.  A  number  of
techniques    has   been   suggested    to   address    these   issues
\cite{Lehner-Reula-Tiglio-2004:multipatch-scalar-field-Kerr-background}:
Multiple   coordinate  patches,   typically  adapted   to  approximate
symmetries  of  certain  solution  domains, combined  with  high-order
operators  are expected to  increase the  accuracy of  any model  of a
stellar system.   Cauchy--perturbative matching provides  an efficient
way to  accurately model the  propagation of gravitational waves  to a
distant  observer,  and  to  yield  physical  boundary  conditions  on
incoming  modes   of  the  Cauchy   evolution.   Constraint-preserving
boundary  conditions  isolate the  incoming  modes  on the  constraint
hypersurface,  and, finally,  for  evolving black  holes, an  excision
boundary is desirable to concentrate  on the behaviour of the external
spacetime.  Only  recently the consideration of  the well-posedness of
the differential  system and the  application of theorems  on discrete
stability  of the  numerical  system  have provided  hints  as how  to
address the  outstanding issues.  In this paper,  we have  applied all
these techniques to a model system: a spherically symmetric black hole
coupled to a massless Klein-Gordon field.

        We find  that the use of a  first-order hyperbolic formulation
of Einstein's field equations, combined with high-order derivative and
dissipation operators with  the summation-by-parts property, penalized
inter-patch   boundary  conditions  and   constraint-preserving  outer
boundary  conditions   leads  to   a  stable  and   accurate  discrete
model. Specifically, isolated Schwarzschild black holes in coordinates
adapted to  the Killing  fields, and in  coordinates on which  a gauge
wave is  imposed, and Schwarzschild black holes  accreting scalar wave
pulses  were taken as  typical model  systems involving  excision. The
results show  that the introduction of several  coordinate patches and
of  a  Cauchy--perturbative  matching  interface  does  not  introduce
significant  artefacts  or  instabilities.  Rather,  the  high-order
methods  allow the  accurate  long-term evolution  of accreting  black
holes  with excision and  Cauchy--perturbative matching  in reasonable
resolutions. As an example, we  have presented the evolution of such a
system with  the high-order SBP operator  $D_{6-5}$, which, at a
resolution of $\Delta x =  M/10$, introduced an error of only $0.3 \%$
after an evolution time of $1,000,000 M$.

        Most system  of interest in  general relativistic astrophysics
will necessarily  require the use of  three-dimensional codes. Results
from a one-dimensional study are useful in that they (i) allow to
gain experience in  a clean but non-trivial physical  system, (ii) can
be easily  reproduced without the need  to implement three-dimensional
codes  with multiple  coordinate patches  and (iii)  allow  to isolate
sources of  difficulty in  the three-dimensional setting  more easily.
With the promising  results from this study, we will,  as a next step,
apply  these techniques to  a three-dimensional,  general relativistic
setting.

\begin{acknowledgements}

We  would  like to  thank  Jorge Pullin, Erik Schnetter and Luis Lehner for
helpful comments.

One  of  the  authors  (B.Z.)  would  like to  thank  the  Center  for
Computation and  Technology at  Louisiana State University  for giving
him the opportunity of an extended visit.

This research was supported in part by NSF under Grants PHY050576 and 
INT0204937 and by NASA under Grant NASA-NAG5-1430 to Louisiana State
University, and employed the resources of the Center
for Computation and Technology at Louisiana State University, which is
supported  by  funding from  the  Louisiana legislature's  Information
Technology  Initiative. 

\end{acknowledgements}


\bibliographystyle{apsrev}

\end{document}